\documentstyle[aps,pre,epsf]{revtex}

\begin{document}
\title{Early-time critical dynamics of lattices of coupled chaotic maps}
\author{Philippe Marcq$^1$ and Hugues Chat\'e$^2$}
\address{$^1$ Department of Physics, Graduate School of Science,
Kyoto University, Kyoto 606, Japan\\
$^2$ CEA --- Service de Physique de l'Etat Condens\'e,
Centre d'Etudes de Saclay, 91191 Gif-sur-Yvette, France
}

\date{September 19, 1997}
\maketitle
\begin{abstract}
The early-time critical dynamics of continuous, Ising-like
phase transitions is studied numerically for two-dimensional 
lattices of coupled chaotic maps. Emphasis is laid on obtaining
accurate estimates of the dynamic critical exponents $\theta'$ 
and $z$. The critical points of five different models are investigated,
varying the mode of update, the coupling, and the local map. 
Our results suggest that the nature of update is a 
relevant parameter for dynamic universality classes of extended 
dynamical systems, generalizing results obtained previously for the static
properties. They also indicate that the universality observed for
the static properties of Ising-like transitions of synchronously-updated
systems does not hold for their dynamic critical properties.
\end{abstract}
 

\section{Introduction}
\label{sec-int}

The last decade has seen considerable experimental, numerical,
and analytical effort aimed at better understanding
the sustained spatio-temporally chaotic regimes of 
large, homogeneous systems kept far-from-equilibrium by 
an external driving force.
In particular, detailed investigation of
a number of hydrodynamical flow regimes,
including convection, shear flow, and crispation
experiments, has led to a wealth of interesting 
insights into the properties of spatio-temporal chaos \cite{CH93}.
Important pending questions concern the status of the 
asymptotic limit of long time and large system size, as well as
the relationship which may exist between 
classical equilibrium statistical mechanics
and possible statistical descriptions of spatio-temporal
chaos in this ``thermodynamic'' limit \cite{CH93,statmech}.

This article focuses on the critical behavior of 
models of spatio-temporal chaos close to second-order-like
phase transitions which occur in the thermodynamic limit. 
Theoretical work \cite{predictions} has suggested that 
phase transitions in generic non-equilibrium systems made up of
locally interacting subunits belong to
the universality class of Model A \cite{Halperin77}, 
for both static and dynamic critical exponents,
provided that the order parameter is a non-conserved, 
scalar quantity. Being based on coarse-grained Langevin descriptions, 
the approach developed in \cite{predictions} overlooks 
the exact nature of microscopic time-evolution. Its conclusion
also relies on the validity of assumptions generally associated with the 
dynamic renormalization group formalism.

Another significant contribution is that of Miller and Huse. In \cite{MH93},
they introduce a simple lattice dynamical system with microscopic 
Ising symmetry (square lattice of locally-coupled, chaotic, odd maps), 
whose salient feature is the presence of a non-equilibrium 
continuous transition qualitatively similar to the ferromagnetic 
critical point of the two-dimensional Ising model. Ising-like 
transitions between spatio-temporally chaotic phases turn out 
to be a fairly common feature of coupled maps lattices (CMLs):
they are observed for a variety of local maps, 
lattice geometries, and update rules 
\cite{OHern95,thesis,ising,Sakaguchi92}. 
However, contrary to the conjecture 
of \cite{predictions} and to early conclusions based on 
simulations of much smaller systems \cite{MH93}, careful 
analysis of finite-size data obtained from
extensive numerical simulations of the Miller-Huse model
shows that the corresponding phase transition does not
belong to the Ising universality class \cite{thesis,ising}.
The measured correlation-length exponent $\nu= 0.89 \pm 0.02$ 
is significantly lower than $\nu_{\rm Ising} = 1$,
while exponent ratios $\beta / \nu$ and $\gamma / \nu$ 
remain in good agreement with Ising values.
Comparison with related models, in particular with 
transitions of sequentially-updated lattice dynamical
systems, further indicates that synchronous update is the
relevant parameter responsible for departure from Ising 
universality: keeping all other features of the Miller-Huse
model unchanged, Ising static exponents are recovered as soon
as sites are updated sequentially \cite{ising}.

Our main objective is to extend previous work
on static critical exponents to the dynamic critical 
properties of the Miller-Huse model.
The dynamic critical exponent $z$, which quantifies the 
algebraic divergence of coherence times at criticality,
is known to be sensitive to parameters otherwise 
irrelevant for static exponents, such as 
the existence or absence of macroscopic quantities 
conserved under time-evolution \cite{Halperin77}. 
In addition, the nature of update is a relevant
parameter for dynamical universality classes
of Ising systems: synchronous update of 
clusters of spins yields distinct, significantly 
lower values of the dynamic exponent $z$ than
is observed for standard sequential or checkerboard update
\cite{Swendsen87}. Ignoring both
the conjecture of \cite{predictions} and the numerical results
of \cite{ising}, one may thus naively expect phase transitions 
of synchronously and sequentially updated CMLs to be 
characterized by different dynamical exponents. 
Here, we want to confirm, for the dynamic properties of 
Ising-like transitions of lattices
of coupled chaotic maps, the relevance of the mode of update already 
discovered in \cite{ising} for their static properties. Similarly,
the static universality class observed for 
synchronously-updated models is revisited from the point of view of
their dynamical properties.

Compared to Ising systems, the measurement of
static critical exponents turns out to be significantly more 
resource-consuming in the case of CMLs, due in particular to the presence of 
unusually large corrections to scaling \cite{ising}. 
Moreover, extracting reliable values of dynamic critical
exponents from direct simulations is a notably difficult task,
even in a priori simpler cases.
Despite much numerical effort, the value of the dynamical 
exponent $z$ of Model A in dimension $d = 2$ remains somewhat 
controversial (see \cite{Ito93} for a 
survey of work done prior to 1993, and \cite{Adler96} for a 
more recent review). The methodology we apply here to phase
transitions of CMLs is based on recent theoretical work by 
Janssen et al., which proves the existence of a new universal regime
in the early critical dynamics of systems starting
from non-equilibrium (e.g. completely disordered)
initial conditions \cite{Janssen89,Janssen92}.
This regime, termed ``initial critical slip'' or 
``universal short-time behavior'' in the literature,
is characterized by a new non-trivial exponent
$\theta'$, unrelated to the usual static and dynamical 
exponents. The dynamical exponents $\theta'$ and $z$ 
can be readily obtained from the initial scaling properties of 
observables of finite-size systems, as shown analytically 
in \cite{Diehl93}, and first implemented
numerically in \cite{Li94}. Unlike standard methods,
this procedure is nearly free from the difficulties
associated with critical slowing down at the transition 
point: useful simulation times ($T \sim 10^2-10^3$) are typically
much shorter than the finite-size coherence time-scale 
$t_L \sim L^z$. Statistical accuracy is ensured by 
ensemble-averaging over a large 
number of independent realizations. Thanks to
high numerical efficiency, good agreement on
the value of critical quantities such as $\theta'$ 
and $d/z - \theta'$  has been already
reached for Model A \cite{Huse,Grassberger,Okano}.
This makes comparison with other systems easier,
and opens the way to an investigation of the relative universality
of $z$ and $\theta'$ which, based on the theoretical work of Janssen et al
\cite{Janssen89,Janssen92}, are expected to depend on
the same relevant parameters. 

This article is organized as follows: 
current understanding of early-time critical 
dynamics is briefly reviewed in Sec.~\ref{sec-met}.
The methodology we follow 
closely parallels that used by Okano et al. for
of the two-dimensional Ising model with
heat-bath and Metropolis algorithm \cite{Okano}.
The same procedure is used throughout, thus
allowing meaningful comparison between exponents
obtained for different CMLs, as well as with exponents of
Model A. First, the dynamic critical properties of the Miller-Huse model,
a lattice dynamical system with synchronous update,
are investigated in Sec.~\ref{sec-mh}. The model
and its phenomenology are introduced in Sec~\ref{sec-mh-model}.
Simulations pertaining to the measure of the
critical exponents $\theta'$ and $z$ are next described
in Sec.~\ref{sec-mh-theta} and \ref{sec-mh-z} respectively.
In Sec.~\ref{sec-update}, we investigate the role played by the type of update
for the dynamic critical properties of Ising-like transitions, 
in order to extend its relevance, already established
in \cite{ising} at the static level.
In Sec.~\ref{sec-update-asyn}, we first consider 
a sequentially updated model introduced in
\cite{ising}, which, according to previous numerical results, 
belongs to the Ising universality class for static critical exponents.
In Sec.~\ref{sec-update-saka}, we turn to 
Sakaguchi's model \cite{Sakaguchi92}, a CML with checkerboard 
update whose static critical exponents are known exactly 
to be equal to those of the Ising model.
Next, we consider, in Sec.\ref{sec-theta'}, various synchronously-updated
models to investigate 
whether the universality of the static critical properties of their
Ising-like transitions extend to their dynamic exponents. 
Our results are summed up and discussed in Sec.~\ref{sec-disc}.

\section{Early-time critical dynamics at second-order transitions}
\label{sec-met}

In order to measure the dynamic critical exponent $z$ from
numerical simulations of finite-size systems, most 
methods considered until a few years ago
made use of the so-called nonlinear
relaxation regime, by, e.g., looking at the
decay of the system's time-dependent magnetization $M(t)$ 
according to:
\begin{equation}
\label{eq-sc-nlrelax}
M(t) \sim t^{-\beta/(\nu z)},
\end{equation}
or similar relations involving higher-order moments.
This regime was generally believed to be relevant within
the time interval $1 \ll t \ll t_L$, whereas finite-size linear 
relaxation eventually prevails beyond $t_L = L^z$,
where the magnetization decays exponentially:
$M(t) \sim \exp(-t/t_L)$.

A point overlooked until the work of
Janssen et al. \cite{Janssen89} is the importance
of initial conditions. Suppose that we start from
disordered non-equilibrium initial conditions
(magnetization is zero or very close to zero)
with very short initial correlation length,
and quench the system to its critical point.
One qualitatively expects fluctuations to be negligible
at first: the system is then mean-field-like.
Since the mean-field critical temperature is usually larger
than the actual critical temperature, the system is in 
its ordered phase, 
and the magnetization (and correlation length) will want
to grow. This accounts for initial magnetization growth.
There is of course a crossover point, after which
the system's behavior reverts to the usual (relaxational) 
behavior of Eq.~\ref{eq-sc-nlrelax}. 

This qualitative idea has been formalized, and the influence
of initial conditions on renormalization-group 
transformations investigated in detail for bulk systems 
\cite{Janssen89,Janssen92}.
Let $m_0$ be the initial magnetization at time $t=0$.
This field gives way to a new scaling index $x_0$ independent
from already known ones (both static and dynamical), and to a 
timescale $t_0$ within which
a new universal scaling regime sets in. The new exponent
$\theta'$ is universal in the same sense as the ususal
dynamic critical exponent $z$, since it was obtained within 
the same formalism. For $t_{\rm mic} \le t \le t_0$, and $m_0$
small enough, the magnetization grows as a power law:
\begin{equation}
\label{eq-sc-theta}
M(t) \sim m_0 t^{\theta'},
\end{equation}
where $\theta' = (x_0 - \beta/\nu)/z$. The microscopic time $t_{\rm mic}$ is the time
after which macroscopically correlated regions form, i.e. regions large
compared to the microscopic length scale, in this case the lattice
constant. The time-evolution of observables for $t \le t_{\rm mic}$
is nonuniversal, and depends on microscopic features of
the model. The crossover time $t_0$ is obtained by 
matching Eqns.~(\ref{eq-sc-nlrelax}) and (\ref{eq-sc-theta}):
\begin{equation}
\label{eq-ics-tslip}
t_0 \sim m_0^{-x_0/z},
\end{equation}
and diverges in the limit of zero initial magnetization
$m_0$. In that case, the nonlinear relaxation regime
is not observed in the bulk.

Then, finite-size scaling theory was introduced by \cite{Diehl93}.
We will need it for interpretation of numerical experiments.
The scale invariant expression reads, for a system of finite size $L$,
and the $k^{\rm th}$ moment of the order parameter, at the critical point:
\begin{equation}
\label{eq-fss-1}
M^{(k)}(t,L,m_0) = b^{-k \beta / \nu} \hat{M}^{(k)}
\left(b^{-z} t, b^{-1} L, b^{x_0} m_0 \right),
\end{equation}
where $b$ is a scaling factor and $\hat{M}^{(k)}$
is a universal function, independent of 
microscopic details of the system.
Important point: the exponent $z$ in Eq.~\ref{eq-fss-1} is the
same as the usual one \cite{Halperin77}.
Choosing the arbitrary prefactor equal to $b \sim t^{1/z}$,
one obtains:
\begin{equation}
\label{eq-fss-2}
M^{(k)}(t,L,m_0) = t^{-k \beta / \nu} \hat{M}^{(k)}
\left(t/t_L, t/t_0 \right),
\end{equation}
where $t_0 = m_0^{-x_0/z}$ and $t_L = L^z$.
For a finite-size system and evolution times $t < t_0, t_L$,
one obtains:
\begin{equation}
\label{eq-mes-dyn-mag}
M(t) \sim m_0 t^{\theta'},
\end{equation}
for small values of $m_0$ \cite{Janssen89}.
This allows to measure $\theta'$ directly.
Then, assuming that the value of $\beta/\nu$ is already
known, $z$ can be obtained thanks to the relation 
\cite{Diehl93}:
\begin{equation}
\label{eq-mes-dyn-m2}
M^{(2)}(t) \sim t^\zeta \;\;{\rm with}\;\; \zeta \equiv (d-2\beta/\nu)/z \,,
\end{equation}
as used for Model A in \cite{Okano}.
Finally, careful renormalization group analysis leads to the 
following scaling form
for the order-parameter time correlation function
$A(t) = \langle m(t) m(0) \rangle$ 
(cf. explicit derivation in \cite{Janssen92})
\begin{equation}
\label{eq-mes-dyn-A}
A(t) \sim t^{-\delta} \;\;{\rm with}\;\; \delta \equiv d/z - \theta' \, ,
\end{equation}
where the space dimension is denoted $d$.

Note that finite-size scaling relations may also be used
in order to measure $z$ and $\beta/\nu$ \cite{Li95}:
\begin{equation}
\label{eq-mes-col1}
\begin{array}{lcl}
U(t,L) &=& U\left(b^z t, bL\right),\\
M^{(2)}(t,L) &=& b^{2 \beta/\nu} M^{(2)}\left(b^z t, bL\right).
\end{array}
\end{equation}
We choose not to, since finite-size effects seem to be either negligible,
or easily controlled in cases relevant here (see below).

To conclude this section, we briefly review recent work on 
the critical dynamics of Model A. The relevant exponent
values are gathered in Table I. The exponent $\theta'$ 
has been measured twice according to 
Eqn.~\ref{eq-mes-dyn-mag}, first by Grassberger
\cite{Grassberger} ($\theta' = 0.191(3)$, heat-bath dynamics), then
by Okano et al. \cite{Okano} ($\theta' = 0.194(4)$, heat-bath and
Metropolis algorithms), thanks to slightly 
different methods. 
We use a conservative combination of the two estimates
as our reference value:
\begin{equation}
\label{eq-modela-theta}
\theta'_{\rm Model A} = 0.193(5).
\end{equation}
Excellent agreement has also been reached
for the combination $\delta = d/z - \theta'$, obtained from 
Eq.~\ref{eq-mes-dyn-A}, between the early measures
of Huse and of Humayun and Bray \cite{Huse} ($\delta = 0.74(1)$,
heat-bath algorithm) and a recent confirmation by Okano et al.
\cite{Okano} ($\delta = 0.739(5)$). Our conservative
estimate is thus:
\begin{equation}
\label{eq-modela-delta}
\delta_{\rm Model A} = 0.74(1).
\end{equation}
The case of the dynamic critical exponent $z$ is more 
delicate. Estimates using methods derived from the theory
of early-time critical dynamics vary between $2.155(3)$ 
(heat-bath) \cite{Okano}, $2.137(11)$
(Metropolis) \cite{Okano}, and $2.143(5)$
(heat-bath) \cite{Li95}. These estimates are somewhat
lower than the currently accepted value $z = 2.165(15)$
\cite{Adler96}, obtained from both series expansions
($z = 2.165(15)$, data from \cite{Damman} reanalized by 
Adler, see \cite{Adler96})
and from a number of direct simulations of very large systems: 
$2.165(10)$ (nonlinear relaxation \cite{Ito93}),
$2.172(6)$ (damage spreading \cite{Grassberger}), 
$2.160(5)$ (non-linear relaxation \cite{Linke}).
Since the latter generally correspond to significantly
better statistics and larger system sizes, we choose
\begin{equation}
\label{eq-modela-z}
z_{\rm Model A} = 2.165(15)
\end{equation}
as our reference value. It leads to
the combination $\zeta = (d-2\beta/\nu)/z = 0.808(6)$,
for $\beta/\nu = 1/8$, in reasonable agreement, within error
bars, with the value obtained
from Eq.~\ref{eq-mes-dyn-m2} in \cite{Okano}:
$\zeta  = 0.817(7)$.

\section{Dynamic critical exponents of the Miller-Huse model}
\label{sec-mh}

\subsection{The model}
\label{sec-mh-model}

Recently, Miller and Huse introduced a CML designed to be a 
simple non-equilibrium Ising-like model \cite{MH93}. 
Its local map, which provides the ``reaction'' part of this reaction-diffusion
lattice dynamical system, 
is an odd, piecewise-linear, chaotic map of the real interval $[-1,1]$:
\begin{equation}
\label{eq-def.hm}
f(x) = \left\{ 
\begin{array}{llclc}
-3x-2 \; &{\rm if} &-1&\le x \le&-{1 \over 3},\\
3x \; & {\rm if} &-{1 \over 3}&\le x \le&{1 \over 3},\\
-3x+2 \; & {\rm if} &{1 \over 3}&\le x \le& 1.
\end{array}
\right.
\end{equation}
The constant absolute value of the slope being equal to three,
its Lyapunov  exponent is positive and equal to $\ln 3$.
For the simple case of a two-dimensional square lattice, the evolution rule
reads:
\begin{equation}
\label{eq-syncupdate}
x_{i,j}^{t+1} = (1-4 g)\; f(x_{i,j}^t) + g \; \left(
f(x_{i-1,j}^{t}) +  f(x_{i,j-1}^{t}) +  f(x_{i+1,j}^t)
+  f(x_{i,j+1}^t) \right),
\end{equation}
where $t$ denotes the (discrete) time, and the subscripts the position on
the lattice. 
The nearest-neighbor coupling constant $g$
can vary between $0$ and $1/4$. In the following, all numerical calculations
are performed on square arrays of linear size $L$ with periodic boundary
conditions.

Since $f(x)$ is an odd function of $x$, discrete spin variables can be defined
in a natural fashion:
\begin{equation}
\label{eq-def.spin}
\sigma^t_{i,j} = \mbox{sign} \left(x_{i,j}^t\right) \in \{-1,1\}.
\end{equation}
Next, the fluctuating, space-averaged magnetization is defined as:
\begin{equation}
\label{eq-mh.mag.inst}
m^t_L = {1 \over L^2} \; \sum_{i,j} \; \sigma_{i,j}^t.
\end{equation}
In fact, one can also use a definition of the ``magnetization'' based 
on the original continuous variables $x^t_{i,j}$. This does not alter 
significantly the statistical results, as we mention in the following.

Increasing the coupling constant $g$, the only control parameter 
in this system, an Ising-like phase transition takes place from
a disordered phase with zero average magnetization at weak coupling
to an ordered phase at strong coupling, where the
spins tend to be aligned with each other. 
The order parameter is the magnetization $M_L = \langle | m^t_L |  \rangle$,
where the brackets represent in practice (long) time-averages
(ergodicity is assumed).

Note that chaos is extensive in
this system \cite{OHern95}, and that
dynamical quantifiers, such as the Kolmogorov-Sinai entropy, 
seem to be insensitive to the onset of long range order at least for the
finite size systems for which these calculations can be made.
Only one lengthscale, the correlation length $\xi$, 
diverges in the thermodynamic limit.
This proves that the transition exists in the thermodynamic 
limit, as corroborated by the applicability of finite-size 
scaling arguments. Such an analysis allows to measure the standard static 
critical exponents, and, in particular, the deviation of 
the correlation-length exponent $\nu = 0.89(2)$,
from the Ising value $\nu=1$. \cite{ising}  

\subsection{Measure of $\theta'$}
\label{sec-mh-theta}

Here we want to look at the short-time dynamics of carefully-prepared
initial configurations with a given initial magnetization $m_0$.
They are generated easily by the following procedure.
For $m_0 = 0$, assign
real random numbers ($x_{i,j}$) uniformly distributed
on $[0,1]$ to $L^2/2$ randomly chosen sites 
of the lattice. Assign then the opposite values ($-x_{i,j}$) randomly
to the remaining $L^2/2$ sites: the total magnetization 
is then exactly zero.
In order to obtain a small, but non-zero initial magnetization,
implement the same procedure, but based on $(L^2-K)/2$ randomly
chosen sites. Then set the value of the $K$ other sites
to, e.g., $x=1$. The magnetization is thus equal to $m_0 = 2 K /L^2$.
We have checked that this particular choice does not
influence the scaling properties described in the following. Choosing $x=1$
possesses the advantage that the initial magnetization has the same value
whether considering discrete spins $\sigma^t_{i,j}$ or the original
continuous variables  $x^t_{i,j}$.

\vspace{1.0cm}
\begin{figure}[thb]
\centerline{
\epsfxsize 5.5cm
\epsffile{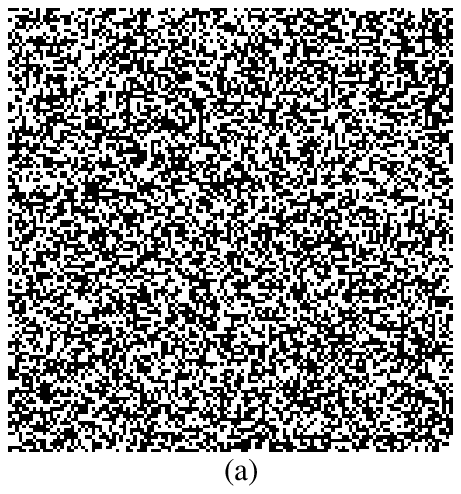}
\hspace{0.5cm}
\epsfxsize 5.5cm
\epsffile{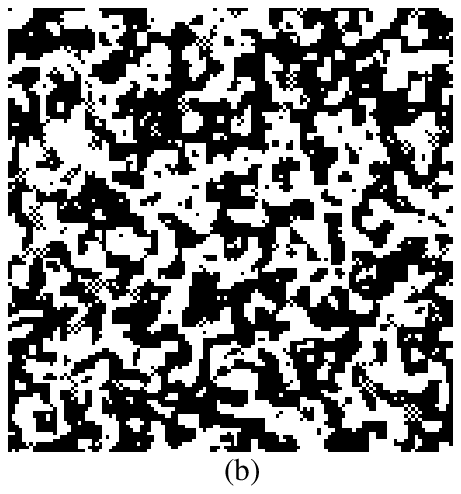}
\hspace{0.5cm}
\epsfxsize 5.5cm
\epsffile{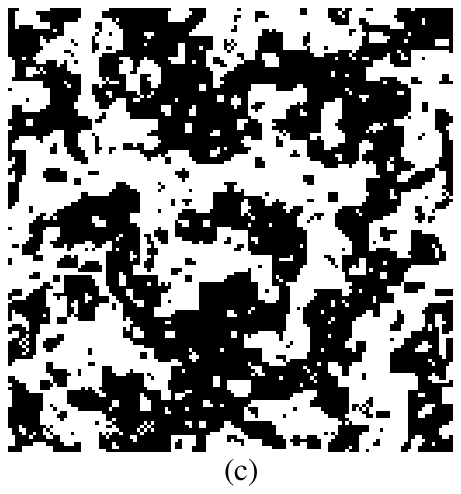}
}
\vspace{0.5cm}
\caption{Snapshots of the coarsening process: positive (resp.
negative) spins are represented by a black (resp. white)
pixel on a two-dimensional grid. This simulation of the Miller-Huse model
is performed at the critical point $g_c = 0.20534$,
for an initial magnetization $m_0 = 2.4\;10^{-4}$
($m_0 = 2 K/L^2, K = 2, L = 128$). Snapshots $(a)$, $(b)$
and $(c)$ correspond to evolution times $t = 0$ 
(completely disordered initial condition), 
$t = t_{\rm mic} = 5$ (build-up of macroscopic 
correlations) and $t = T = 128$ (``initial critical slip''
regime).
}
\label{fig-mh4.snap}
\end{figure}

\begin{figure}[thb]
\centerline{
\epsfxsize 10cm
\epsffile{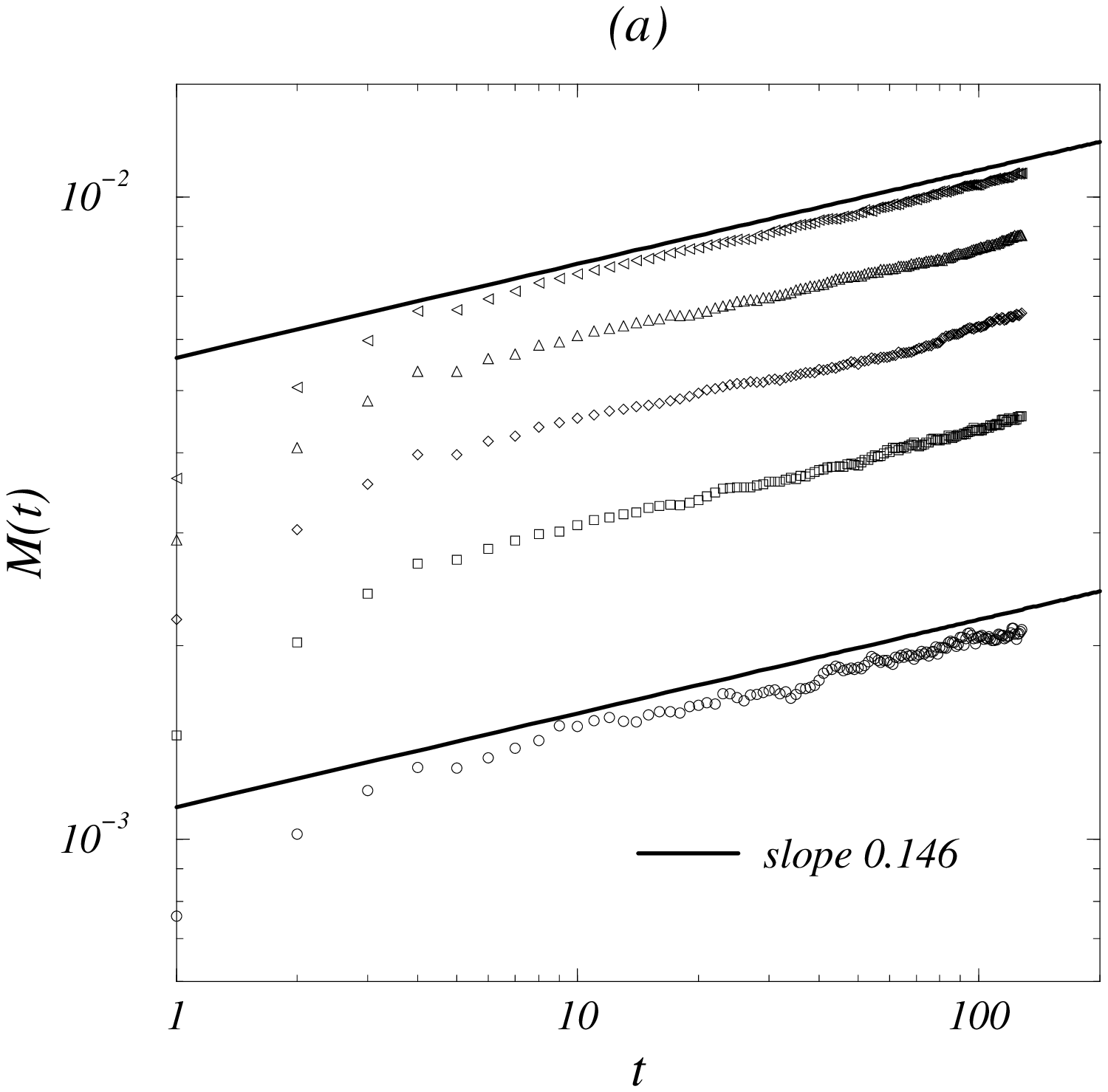}
\hspace{-1.5cm}
\epsfxsize 10cm
\epsffile{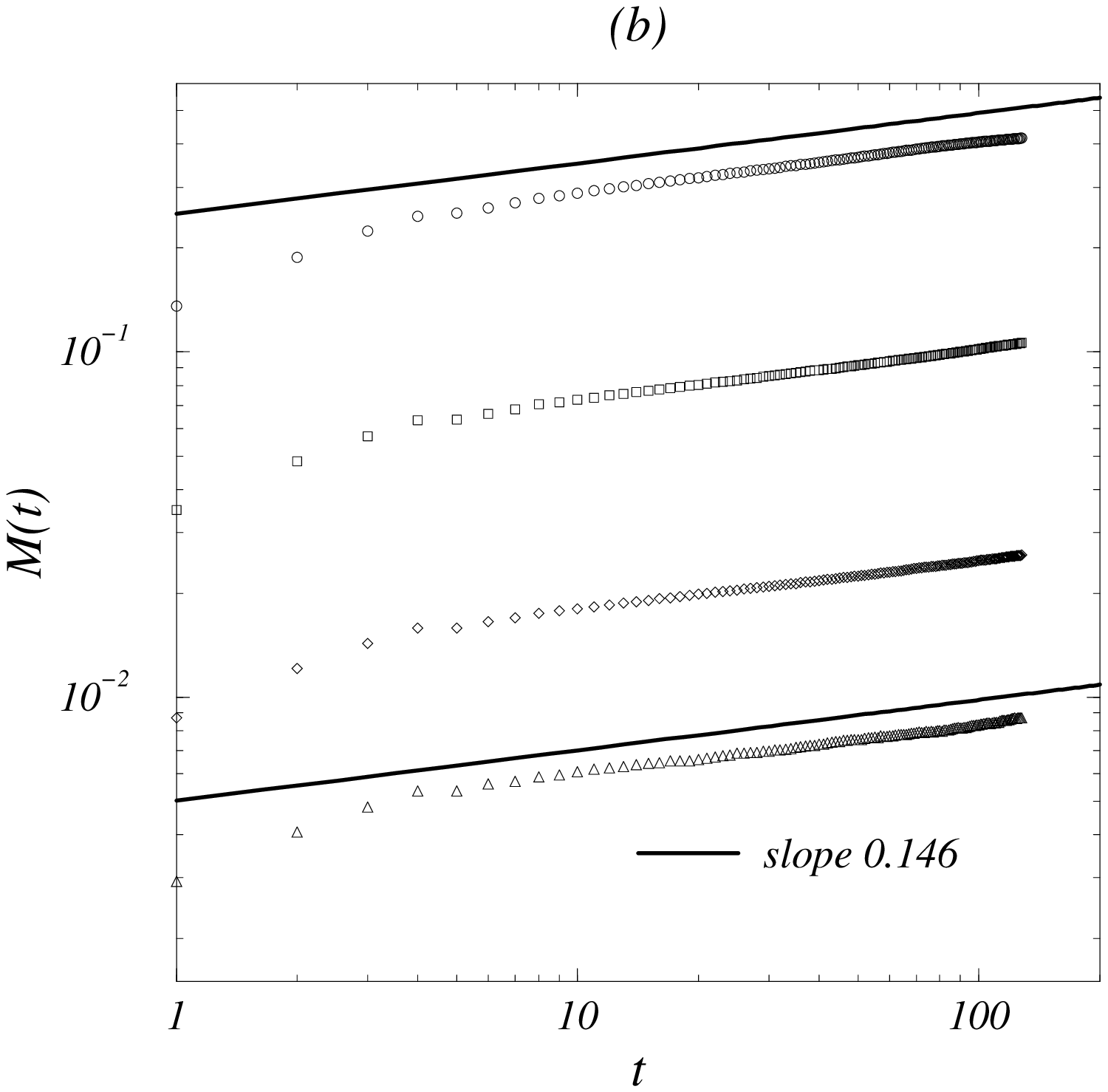}
}
\vspace{-0.5cm}
\caption{Measure of the dynamic critical exponent
$\theta'$ for the Miller-Huse model.
We plot in log-log scale the magnetization $M(t)$ vs. time $t$
measured at the critical point $g = g_c = 0.20534$.
Clear scaling is observed in all cases for 
$t \ge t_{\rm mic} = 5$.
The solid lines in both graphs correspond to a slope equal 
to $\theta' = 0.146$.
Graph $(a)$: the system size $L = 128$ is fixed,
the initial magnetization $m_0 = 2 K /L^2$ 
varies between $m_0 = 2.4\;10^{-4}$ and $1.2\;10^{-3}$,
for $5$ distinct values of $K = 2,4,6,8,10$ (from bottom to top).
Good agreement between slopes corresponding to
different values of $K$ suggests that the limit 
$m_0 \rightarrow 0$ is satisfactorily approximated.
Graph $(b)$: the initial condition $K = 6$ is fixed, for 
system sizes $L = 16, 32, 64, 128$ (from top to bottom).
Finite-size effects are negligible. This suggests
that the infinite-size limit is satisfactorily approximated.}
\label{fig-mh4.theta}
\end{figure}

The value of the critical coupling strength $g_c$ was 
previously obtained according to Binder's method.
We use $g_c = 0.20534(2)$ \cite{ising}.
As predicted in \cite{Janssen89}, a regime
of initial growth of the magnetization is
observed, as well as the crossover toward
nonlinear relaxation for large enough initial
magnetization $m_0$. 
The corresponding coarsening process is 
illustrated in Fig.~\ref{fig-mh4.snap}.
For measurement purposes,
we use $K=2,4,6,8,10$, for sizes ranging between
$L=16$ and $L=128$. The duration of a run is $T = 128$.
In such conditions, no crossover is observed
to the nonlinear relaxation regime, since $t_0 \gg T$.
Thanks to the good quality of our data, the value
of the microscopic time $t_{\rm mic} = 5$ can 
be obtained by simple visual inspection (Fig.~\ref{fig-mh4.theta}).
This relatively small value is comparable to
what has been observed for the two-dimensional 
Ising model \cite{Okano}.
Scaling of the magnetization versus time 
is observed over the time interval $5 \le t \le 128$. 
This corresponds to the initial critical slip regime.

The exponent $\theta'$ is measured thanks to a linear
fit in log-log scale over the interval $t_{\rm mic} \le t \le T$.
We checked that using larger values of $t_{\rm mic}$ and/or $T$
does not alter the estimate. 
Note also that using values of the critical coupling 
outside the confidence interval $g_c = 0.20534(2)$
does not lead to an improved quality of fits:
this confirms the validity of estimates
of the critical coupling obtained in \cite{ising}.

Ensemble averages are performed
over $512000$ realizations for $L \le 64$,
$128000$ realizations for $L =128$.
Statistical errors are estimated by comparing the
exponent values obtained for five different
initial magnetizations $m_0 = 2 K /L^2$,
$K=2,4,6,8,10$. Error bars take into account
the uncertainty on $g_c$. Note that the corresponding values
of $m_0$ are much smaller than those used by Okano 
et al., who needed to extrapolate exponent values
obtained for small, but finite initial magnetization
to the limit $m_0 = 0$. Our procedure is similar
to that used by Grassberger \cite{Grassberger},
since no extrapolation is needed. 
The exponent values thus measured for
$L = 32, 64, 128$ are respectively
$\theta' = 0.148(2), 0.142(4)$ and $0.148(7)$. 
Statistically equivalent values are obtained when considering the
magnetization based on the continuous variables.
Within error bars, no finite-size effect is observed 
for $L \ge 32$. Our global (conservative) estimate is:
\begin{equation}
\label{eq-mh4-theta}
\theta'_{\rm MH} = 0.146(9).
\end{equation}
Note that this result is {\it not} consistent with the accepted
value for the critical dynamics of Model A: 
$\theta' = 0.193(6)$
obtained by similar methods and with a similar statistical
quality in \cite{Grassberger,Okano}.

\subsection{Measure of $z$}
\label{sec-mh-z}

\begin{figure}[thb]
\centerline{
\epsfxsize 10cm
\epsffile{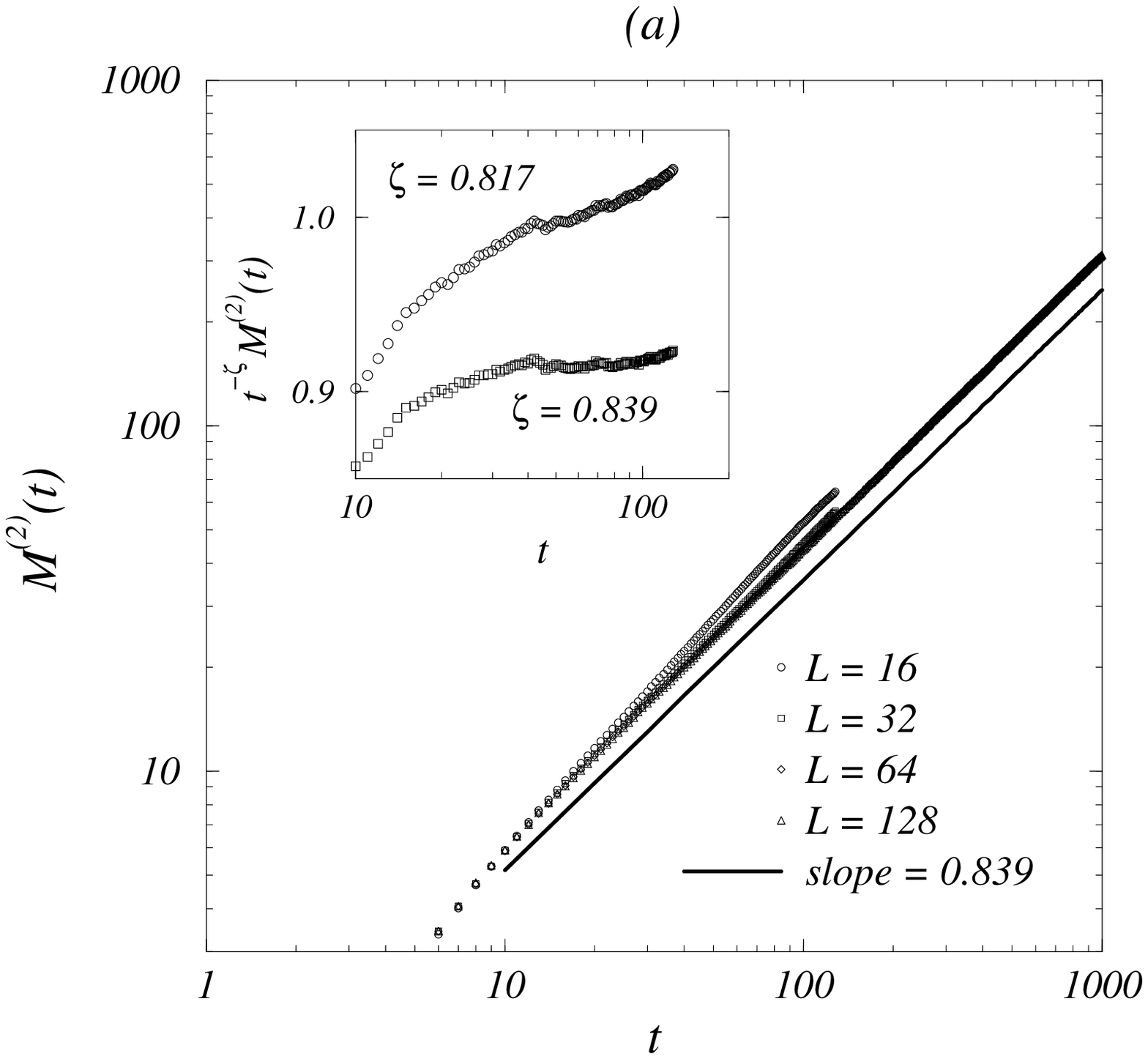}
\hspace{-1.5cm}
\epsfxsize 10cm
\epsffile{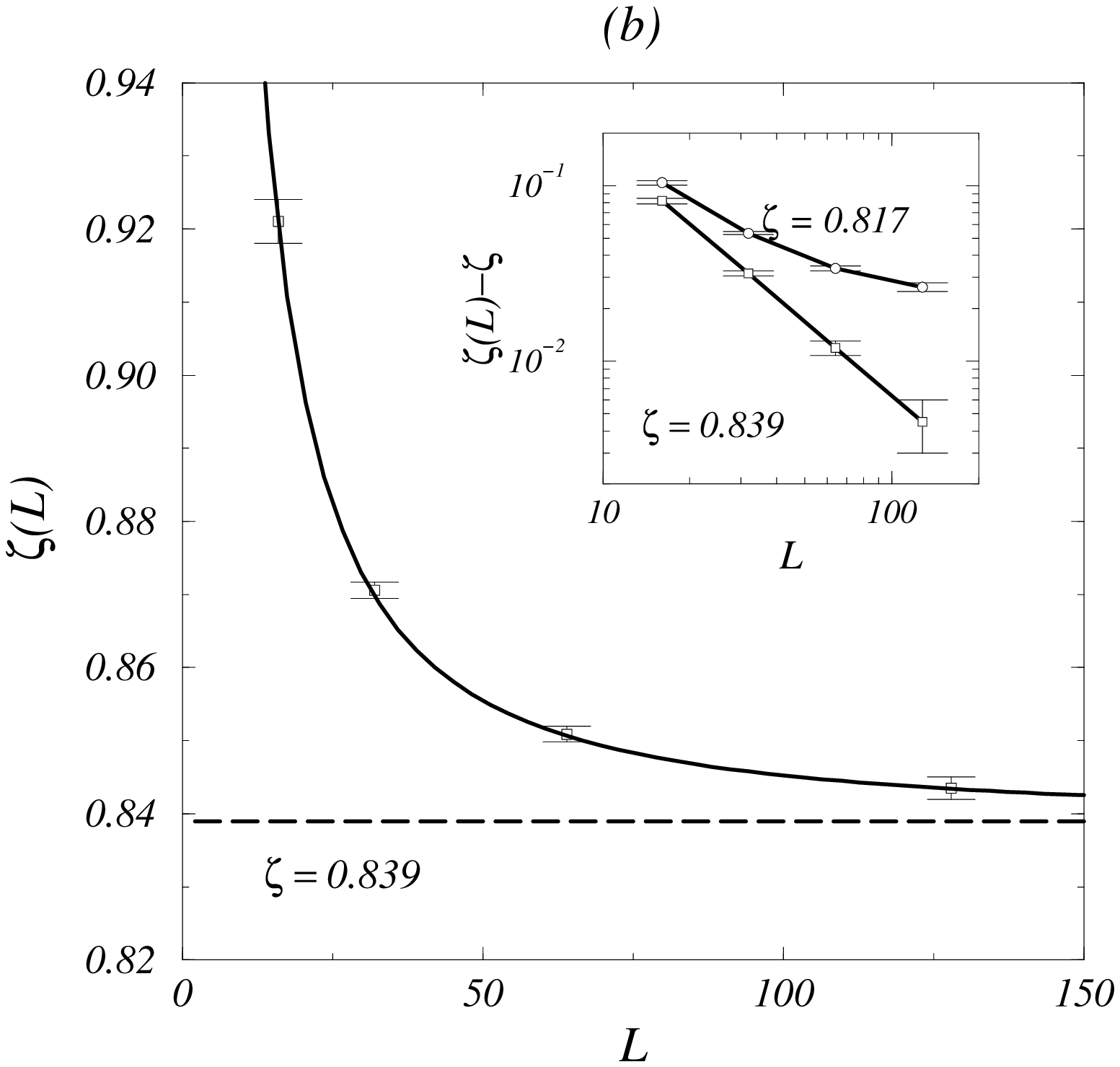}
}
\vspace{-0.5cm}
\caption{Measure of the dynamic critical exponent
$z$ for the Miller-Huse model. 
Graph $(a)$: we plot in log-log scale the second moment 
$M^{(2)}(t)$ vs. time $t$ at the critical point $g = g_c = 0.20534$,
for an initial magnetization $m_0 = 0$.
System sizes are $L = 16, 32, 64, 128$ (from top to bottom). 
Scaling is observed for $t \ge t_{\rm mic} \sim 30$.
The solid line corresponds to a slope equal to 
$\zeta = (d-2\beta/\nu)/z = 0.839$.
The curves corresponding to $L = 64$ and $L=128$ ($T = 128$) are 
undistinguishable. This suggests that finite-size effects
become small for $L \ge 64$. Data obtained for $L=128$,
$T=1024$, $32000$ realizations (triangular symbols)
shows no evidence of cross-over to a distinct time-asymptotic regime.
Insert: we plot the ratio
$M^{(2)}(t)/t^{\zeta}$ vs. time $t$ in log-lin scale
for the values $\zeta = 0.817$
(Model A, top curve) and $\zeta = 0.839$
(best fit for extrapolated value in the infinite-size limit (see (b)), 
bottom curve). 
The system size is $L=128$, $T = 128$. This graph
shows that our (finite-size) data is {\it not} compatible with
exponent values expected for Model A.
Graph $(b)$: we plot the finite-size exponent $\zeta(L)$ 
vs. system size $L$, as obtained from linear fits
of the data presented in Graph $(a)$ for $30 \le t \le 128$.
The infinite-size estimate $\zeta = \zeta(\infty) = 0.839$
(dashed line) is derived from the relation 
$\zeta(L) = 0.839 +  3.9 L^{-1.4}$ (solid line).
The insert shows a log-log plot of $\zeta(L) - \zeta(\infty)$
vs. system size $L$ for the two values 
$\zeta = 0.817$ (Model A) and $\zeta = 0.839$ (measured).
The latter value optimizes the quality of a linear fit of
$\log(\zeta(L) - \zeta)$ vs. $\log(L)$. This plot indicates that 
our data is {\it not} compatible with exponent values expected for Model A
when extrapolated to the infinite-size limit.}
\label{fig-mh4.khi}
\end{figure}

A first way to measure exponent $z$ is given by Eq.~\ref{eq-mes-dyn-m2}.
The experimental conditions are similar to those mentioned
in the last section, but for an initial magnetization equal to zero
($m_0 = 0$, $K=0$). Simulations were performed for
$L= 16, 32, 64, 128$, $T = 128$, ensemble averages performed
over $512000$ realizations for $L \le 64$ and
$128000$ realizations for $L=128$. 
Log-log plots of the second moment $M^{(2)}$ vs time show 
very good scaling, and lead to estimates of the combination
of exponents $\zeta = (d-2\beta/\nu)/z = 7/(4z)$, 
assuming that $\beta/\nu = 1/8$ for $d=2$, as implied by 
\cite{ising}. In order to avoid interferences with the
current experimental uncertainty on $\beta/\nu$
---estimated to be $\beta/\nu = 0.131(6)$ in \cite{ising}---,
we will work with $\zeta$, and convert into $z$ as
late as possible. Longer runs (up to $T=1024$) were performed for
large system sizes ($L = 128$), with lesser statistical accuracy.
This allowed to check that the exponents measured do indeed
correspond to the asymptotic regime (cf. Fig.~\ref{fig-mh4.khi}).

Here, determining the microscopic time $t_{\rm mic}$ 
requires additional effort, when compared to the 
previous case. We use a method introduced by 
Okano et al. \cite{Okano}. Local exponents $\zeta(t)$ 
are first measured from ``local'' fits limited
to an interval of time $[t, t+t_{\rm loc}]$. The microscopic
time $t_{\rm mic}$ is defined as the time beyond which $\zeta(t)$
becomes stationary, within statistical fluctuations.
We find $t_{\rm mic} \sim 30$, for $t_{\rm loc} = 15$ and all sizes $L$ considered. 
This value does not depend on the value of $t_{\rm loc}$
for $t_{\rm loc}$ large enough.
The microscopic time measured here is much larger than 
the one estimated for the scaling of $M(t)$ at small $m_0$, 
as had already been observed for the Ising model \cite{Okano}. 

Asymptotic values of $\zeta$ are then obtained, 
as before, from a global linear fit performed in
log-log scale over the interval $t_{\rm mic} \le t \le T = 128$.
As before, we checked that values obtained do not change, within
error bars, for $15 \le t_{\rm mic} \le 50$. Again, these values were independent
on whether the discrete spins or the continuous variables were used to
calculate the magnetization.

In this case, finite-size effects are sizeable and can be well controlled:
we obtain values of $\zeta(L)$
for sizes $16 \le L \le 128$ which are monotonously decreasing
and seem to converge. Statistical errors are
evaluated from a comparison of  exponent values measured
for three coupling strength values in the confidence
interval $[0.20532, 0.20536]$. As before, the quality of
fits does not increase when varying $g_c$ outside this
interval. In order to evaluate the rate of convergence, 
we use the Ansatz:
\begin{equation}
\label{eq-exp-khi-ansatz}
\zeta(L) - \zeta(\infty) \sim L^{-\omega},
\end{equation}
where $\zeta(\infty)$ is the desired infinite-size exponent.
Although algebraic relaxation toward an asymptotic value
seems natural in the context of critical phenomena,
we are not aware of any theoretical justification 
for Eq.~\ref{eq-exp-khi-ansatz}.
The validity of this phenomenological Ansatz is confirmed by
our data (Fig.~\ref{fig-mh4.khi}), which is
not compatible with, e.g., exponential relaxation. 
Optimizing linear fits in log-log scale
of $\zeta(L) - \zeta(\infty)$ vs. $L$
yields the following estimate: $\zeta(\infty) = 0.839(3)$
(and, incidentally, $\omega = 1.4$).
The Ising value $0.817$ is not compatible with our data
plus Ansatz~\ref{eq-exp-khi-ansatz} (see inserts of Fig.~\ref{fig-mh4.khi}).
Note that we use here the numerical results of \cite{Okano} 
as a reference value for Model A. The discrepancy between
our data for the Miller-Huse model and the currently accepted value 
\cite{Adler96} for Model A ($\zeta = 0.808(6)$, see discussion 
in Sec.~\ref{sec-met}) is even larger.

Using the theoretical, exact Ising value for $\beta/\nu$ \cite{ising},
our estimate yields $z = 2.082(4)$, at variance with the Model A exponent,
irrespective of the method used to estimate it, 
whether it is initial critical slip \cite{Okano,Li95} or 
standard methods \cite{Adler96}.
Using the measured value $\beta/\nu = 0.131(6)$ \cite{ising}, one obtains:
\begin{equation}
\label{eq-exp-khi-value}
z_{\rm MH} = 2.07(2),
\end{equation}
a conservative estimate which we endorse. 

Confirmation of the previously measured values may in principle
be obtained from the scaling behavior of the correlation
function (Eq.~\ref{eq-mes-dyn-A}), with an exponent $\delta = d/z - \theta'$.
Previously measured values lead to $d/z - \theta' = 0.81(1)$.
Even though the runs used are the same as for the second moment $M^{(2)}$,
in practice, scaling is not satisfactory. In particular,
local exponents $\delta_{\rm loc}(t)$ do not converge to
stationary values. Very strong finite-size corrections 
to the dominant scaling are present, which preclude any effective
measurement of $\delta$. However, the observed behavior is compatible
with large-size, long-time convergence to the above value 
$\delta = d/z - \theta' = 0.81(1)$. Direct, reliable estimates for $\delta$
remain beyond our numerical resources.

In conclusion, the above study of the Miller-Huse model first shows
that the regime of initial
critical slip exists also for (some) deterministic systems,
with the same phenomenology as for spin systems.
Quantitatively, our simulations lead to the conclusion that
the Miller-Huse model does not belong to the dynamic
universality class of Model A. 
Since the mode of update is the relevant parameter explaining
departure from the Ising universality for static critical
exponents in this model \cite{ising}, 
one would naturally like to know whether or
not this is also true for dynamic exponents.
Section \ref{sec-update} deals with this question, with the study
of two models with respectively sequential and checkerboard update.

\section{Update rules}
\label{sec-update}

\subsection{Sequential update}
\label{sec-update-asyn}

The model considered here is identical to the Miller-Huse model
except for one point: the update rule.
Evolution rule \ref{eq-syncupdate} is replaced by:
\begin{equation}
\label{eq-asyncupdate}
x_{i,j}^{t+1} = (1-4 g)\; f(x_{i,j}^t) + g \; \left(
f(x_{i-1,j}^{t+1}) +  f(x_{i,j-1}^{t+1} +  f(x_{i+1,j}^t)
+  f(x_{i,j+1}^t) \right).
\end{equation}
Sites are updated one at a time,
in sequential order from the top, leftmost site to the 
bottom, rightmost one, as in:
\begin{equation}
\label{eq-def.succ}
\begin{array}{rl}
\ldots \rightarrow & (1,1) \rightarrow (2,1) \rightarrow (3,1) 
\rightarrow \ldots \rightarrow (L,1) \rightarrow\\
\rightarrow & (1,2) \rightarrow (2,2) \rightarrow (3,2) \rightarrow \ldots 
\rightarrow (L,2) \rightarrow\\
&\ldots\\
\rightarrow & (1,L) \rightarrow (2,L) \rightarrow (3,L) \rightarrow \ldots 
\rightarrow (L,L) \rightarrow \ldots
\end{array}
\end{equation}
where arrows indicate the order of update
between sites of indices $(i,j)$. Consequently, boundary conditions are
helical in this case. (Note that this mode of update 
was termed ``asynchronous'' in \cite{ising}.) 
A continuous transition similar to the Ising ferromagnetic point
occurs in this system too, albeit for significantly lower
coupling strength $g_c = 0.11255(5)$, as measured in \cite{ising}
thanks to Binder's method. Its (measured) static critical exponents
are compatible with the static Ising universality class
$\beta/\nu = 0.117(12)$, $\nu = 1.02(7)$ \cite{ising}.

\begin{figure}[thb]
\centerline{
\epsfxsize 10cm
\epsffile{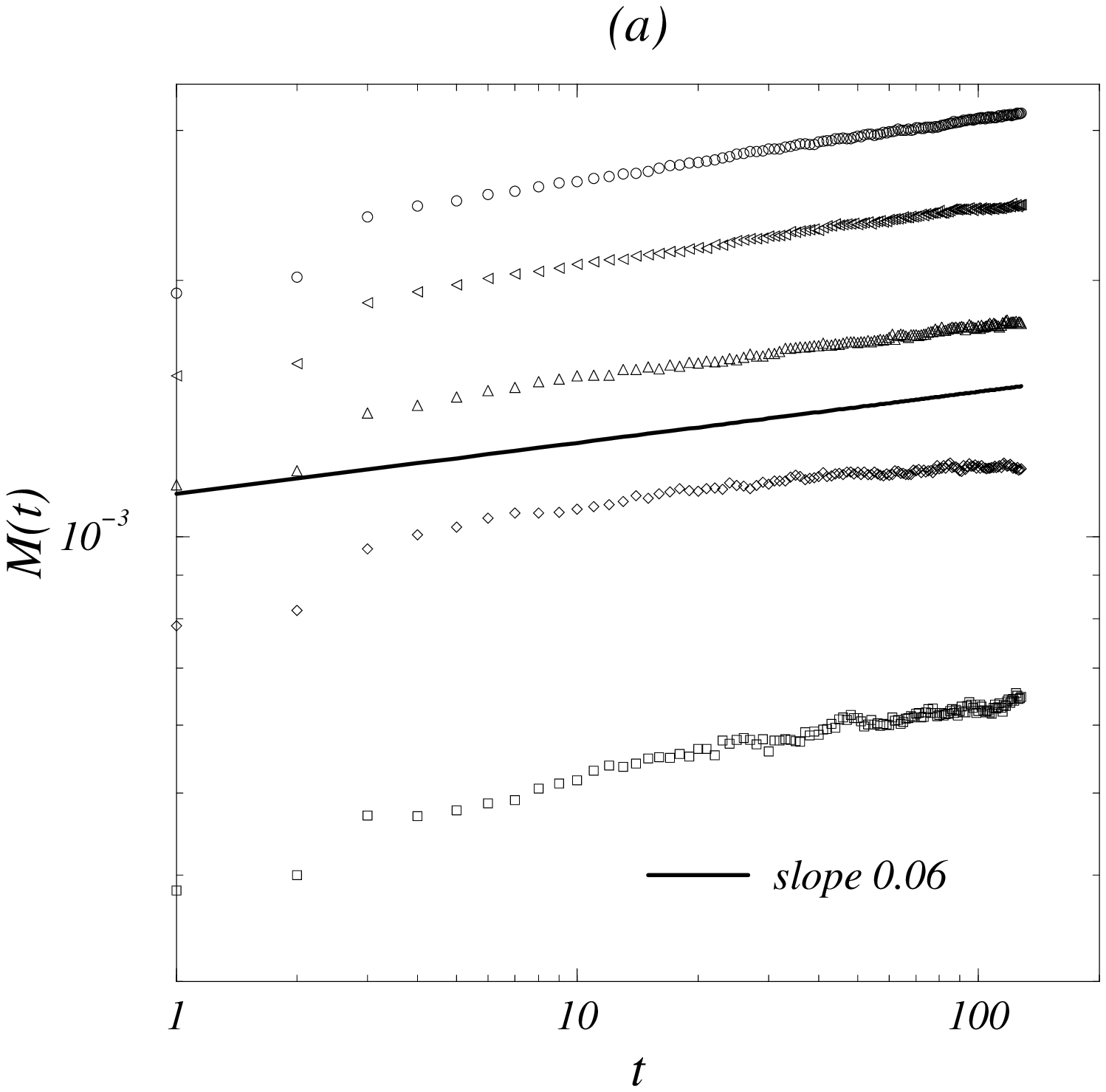}
\hspace{-1.5cm}
\epsfxsize 10cm
\epsffile{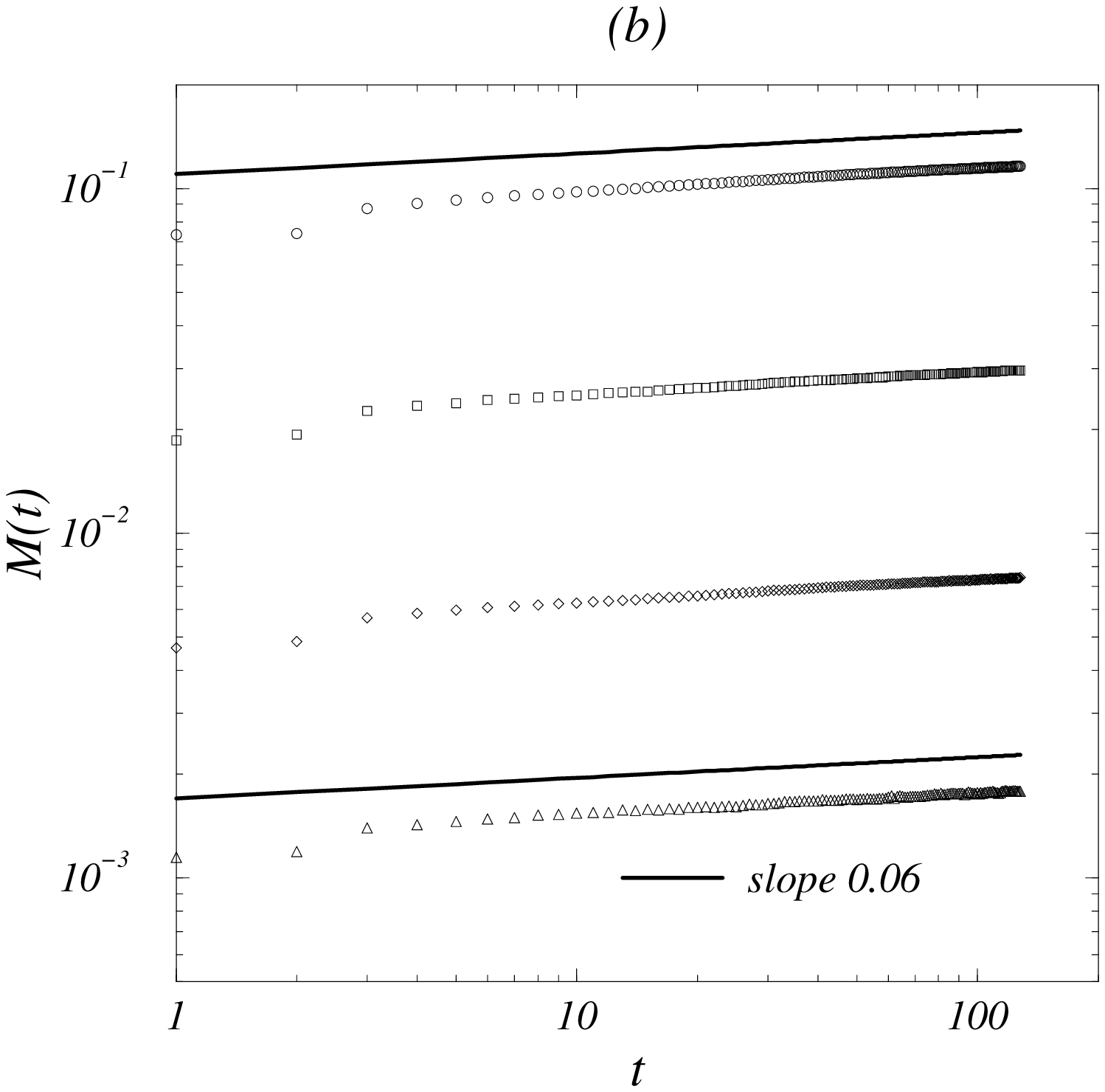}
}
\vspace{-0.5cm}
\caption{Measure of the dynamic critical exponent
$\theta'$ in the case of sequential update.
The magnetization $M(t)$, measured at the critical 
point $g = g_c = 0.11255$, is plotted vs. time $t$ 
in log-log scale. Scaling is observed in all cases
for $t \ge t_{\rm mic} = 5$.
The solid line in both graphs corresponds to a 
slope equal to $\theta' = 0.06$. 
Graph $(a)$: the system size $L = 128$
is fixed, the initial magnetization $m_0 = 2 K /L^2$ 
varies between $m_0 = 2.4\;10^{-4}$ and $1.2\;10^{-3}$,
for values of $K = 2,4,6,8,10$ (from bottom to top).
Graph $(b)$: the initial condition $K = 6$ is fixed,
for system sizes $L = 16, 32, 64, 128$ (from top
to bottom). Finite-size effects are negligible.
}
\label{fig-a4.mag}
\end{figure}

The initial conditions to measure the early-time critical
properties of this transition are prepared as for the Miller-Huse model.
The observed phenomenology is the same, for a comparable
microscopic time $t_{\rm mic} \sim 5$, as estimated visually 
(Fig.~\ref{fig-a4.mag}). The value of
$t_{\rm mic}$ is confirmed by the method described in Sec.~\ref{sec-mh-z}
(local exponent).
Finite-size effects are negligible for $L \ge 32$.
Measurement of the dynamic critical exponent $\theta'$
is based on sizes $L = 32, 64, 128$. The number of
realizations over which ensemble-averaging is performed
is the same as before, $512000$ for $L=32, 64$, and $128000$
for $L=128$ (see Fig.~\ref{fig-a4.mag}). 
Our final estimate is 
\begin{equation}
\label{eq-seq-theta}
\theta'_{\rm Sequential} = 0.06(2).
\end{equation}
Error bars take into account both statistical errors and 
the uncertainty due to the error bars on the location
of the critical coupling strength. This is clearly 
not compatible with either the value obtained for the Miller-Huse
model ($\theta' = 0.146(9)$) or with the Model A
value ($\theta' = 0.193(5)$). 
At this stage we have no particular understanding of the low value of
$\theta'$ in this case. Finally, note that our data is equally well
fitted by a logarithmic time-dependence: $M(t) \sim \log(t)$,
for which no theoretical justification is at present available.

\begin{figure}[thb]
\centerline{
\epsfxsize 10cm
\epsffile{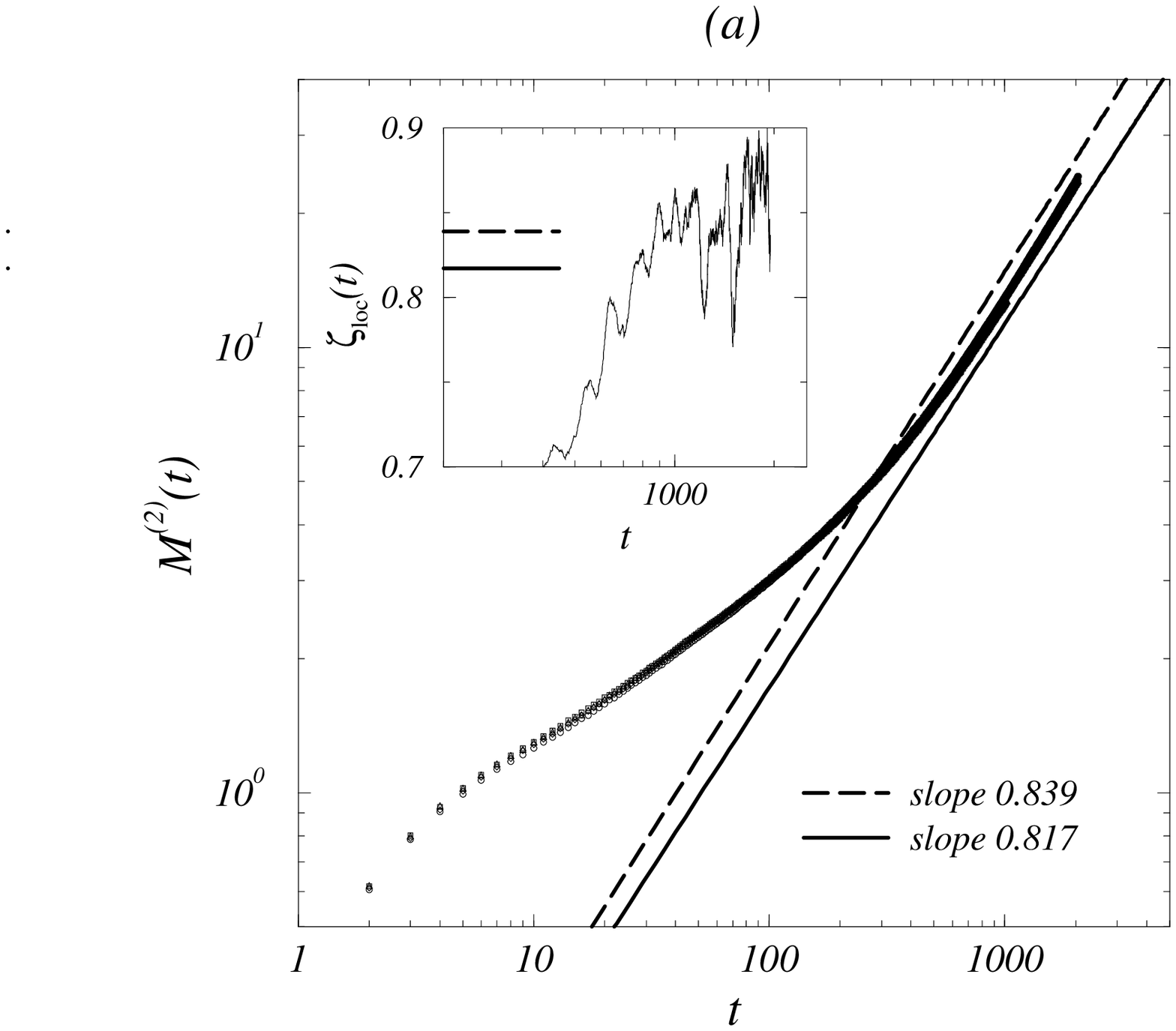}
\hspace{-1.5cm}
\epsfxsize 10cm
\epsffile{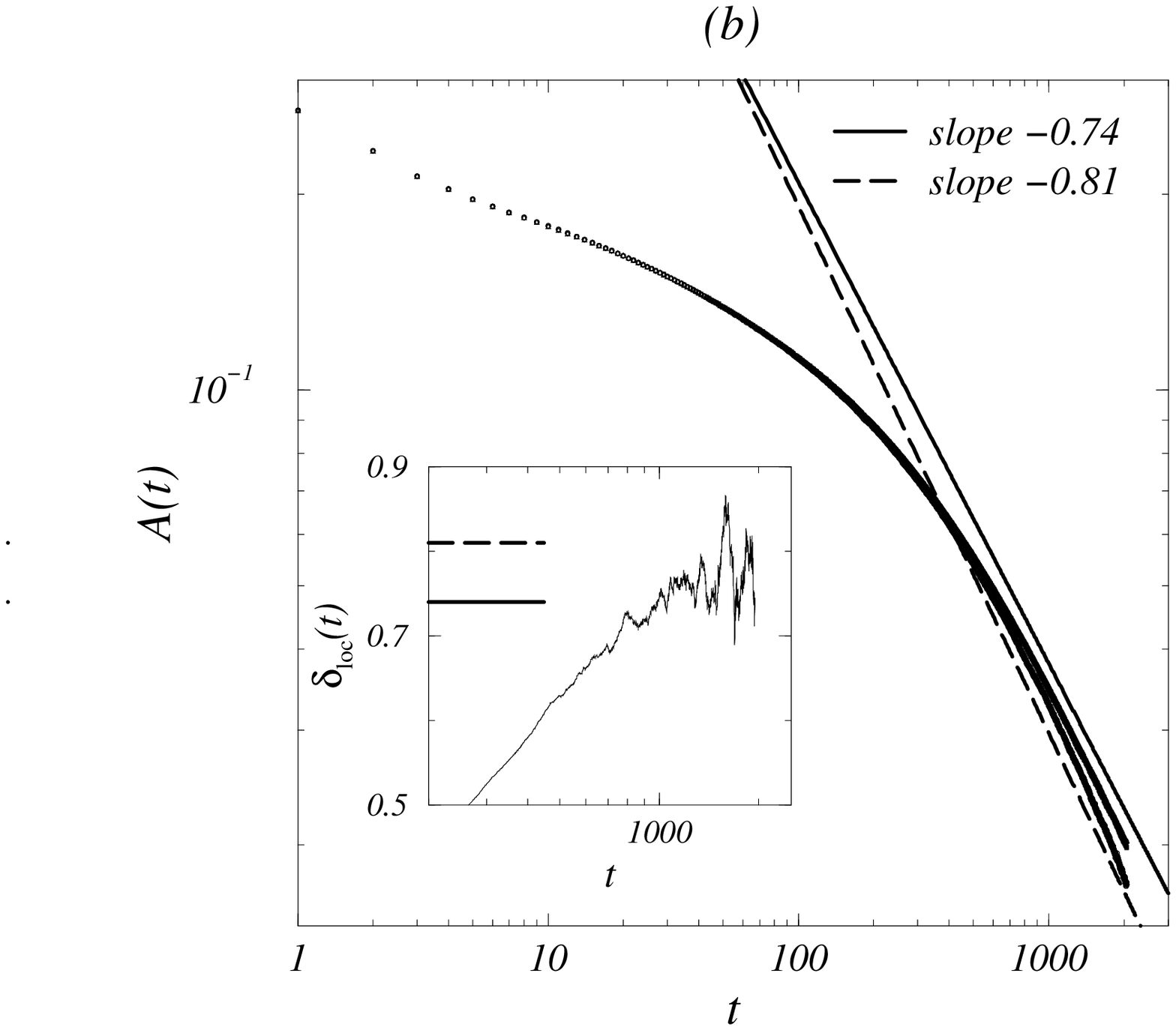}
}
\vspace{-0.5cm}
\caption{Measure of the dynamic critical exponent $z$
in the case of sequential update. The second moment
of the magnetization $M^{(2)}(t)$ (graph $(a)$) and the temporal
autocorrelation function $A(t)$  (graph $(b)$) are plotted
in log-log scale vs. time $t$, as measured at the 
critical point $g = 0.11255$, 
for zero initial magnetization $m_0 = 0$, up to
time $T = 2048$. The system sizes 
considered are $L = 16, 32, 64, 128$.
All curves superpose for $M^{(2)}(t)$, and the last three for $A(t)$. 
The microscopic time $t_{\rm mic}$ can be roughly evaluated to
$t_{\rm mic} = {\cal O}(1000)$, especially from the inserts, where we plot
 in log-lin scale the local
exponents $\zeta_{\rm loc}(t)$ and $\delta_{\rm loc}(t)$ vs. time $t$,
computed for $L=64$, over time intervals of duration $t_{\rm loc} = 100$.
Graph $(a)$: the $\zeta$ values corresponding to Model A ($\zeta=0.817$, 
solid lines) and the Miller-Huse model  ($\zeta=0.839$, dashed lines)
are indicated.
Graph $(b)$: the bottom and top curves respectively 
correspond to sizes $L = 16$ and $L=32, 64, 128$ 
(superposed). The $\delta$ values corresponding to Model A ($\delta=0.74$, 
solid lines) and the Miller-Huse model  ($\zeta=0.81$, dashed lines)
are indicated.
Note that fluctuations
of $\zeta_{\rm loc}(t)$ and  $\delta_{\rm loc}(t)$ observed for 
$t \ge t_{\rm mic}$ encompass the values for both models.
}
\label{fig-a4.dyn}
\end{figure}

Measuring the exponent $z$ is much more difficult, since 
the microscopic time (or rather the time beyond which 
corrections to dominant scaling vanish and the asymptotic regime
sets in) is very large, of the order of $t_{\rm mic} \sim 1000$. 
This is clearly shown in Fig.\ref{fig-a4.dyn},
and in particular from the plots of the local exponents
vs. time in the inserts. Ensemble averages are computed over
typically $64000$ independent runs, for a simulation time
of $T = 2048$. Estimates of exponents must be derived from
intervals much smaller than one decade (typically 
$1000 \le t \le 2048$). Linear fitting in log-log scale is
thus impractical. For that reason, we assume that the plateau
observed for $t > t_{\rm mic}$ in plots of the local exponents
correspond to the asymptotic value, and evaluate error bars
from the variation of local exponent within that interval 
(Fig.\ref{fig-a4.dyn}). We obtain: 
\begin{equation}
\label{eq-seq-z-delta}
\begin{array}{lcl}
\zeta_{\rm Sequential} & = & 0.83(6),\\
\delta_{\rm Sequential} & = & 0.78(8).
\end{array}
\end{equation}
The estimated value of $\zeta$, close to that of the Miller-Huse model,
leads to $z = 2.12(15)$ (assuming $\beta/\nu = 1/8$),
with error bars large enough to include the uncertainty
on the exponent of both Model A and Miller-Huse model.
The estimated value of $\delta$ is closer to that of Model A, but 
is too imprecise to be exploited.

We are thus unable to give an estimate of
$z$ which would decide between the values for the Model A, the Miller-Huse
model, or an eventual third number. It is clear, though, that our
estimate of exponent $\theta'$ ($\theta'=0.06(2)$) is not compatible 
with the values of either Model A  ($\theta'=0.193(5)$) 
or the Miller-Huse model  ($\theta'=0.146(9)$).

The above results confirm, at the dynamic level, 
those obtained in \cite{ising} for the static
critical properties of Ising-like phase transitions in systems made of
coupled chaotic maps: the mode of update is relevant. 
In \cite{ising}, the static exponents of the sequential-update model 
studied above were measured to be, within numerical accuracy, those
of the Ising model. The above results show that these two models have different
dynamic exponents (at least $\theta'$). Is this due to their mode of update
and/or to other factors?
In the next subsection, we introduce Sakaguchi's model, a system of 
coupled chaotic maps designed to have exactly the static exponents of 
the Ising model, in order to further assess the relevance of the mode
of update and that of the nature of the model for the dynamic 
scaling properties of Ising-like transitions.

\subsection{Checkerboard update}
\label{sec-update-saka}

A few years ago, Sakaguchi introduced a system of coupled Bernoulli
maps  with an exponential coupling scheme and checkerboard update,
which leads exactly to the Ising equilibrium Gibbs measure
\cite{Sakaguchi92}. 

Two continuous  variables,
$x_{i,j}^t$ and $\Delta_{i,j}^t$ in $[-1, 1]$, 
are defined on each site of a two-dimensional square lattice with periodic
boundary conditions. Their evolution rule reads:
\begin{equation}
\label{eq-saka-update}
\begin{array}{lclll}
x_{i,j}^{t+1} &=&  {2 \over 1 + \Delta_{i,j}^t} 
\left( x_{i,j}^t + 1 \right) -1, & \; -1 <  x_{i,j}^t <  \Delta_{i,j}^t,\\
x_{i,j}^{t+1} &=&  {2 \over 1 - \Delta_{i,j}^t} 
\left( x_{i,j}^t - 1 \right) +1, & \; \Delta_{i,j}^t <  x_{i,j}^t <  1,
\end{array}
\end{equation}
where the (time-dependent) slopes $\Delta_{i,j}^t$ of the Bernoulli maps 
are calculated according to:
\begin{equation}
\label{eq-def.slope.saka}
\Delta^t_{i,j} = \mbox{tanh} \left\{ J \left(
\sigma_{i-1,j}^{t-1} + \sigma_{i+1,j}^{t-1} + 
\sigma_{i,j-1}^{t-1} + \sigma_{i,j+1}^{t-1} 
\right) \right\},
\end{equation}
with $J$ a coupling constant. 
Discrete spin variables can be defined by:
\begin{equation}
\label{eq-def.spin.saka}
\sigma^t_{i,j} = \mbox{sign} \left(x_{i,j}^{t+1} - x_{i,j}^t \right) 
\in \{-1,1\} \,,
\end{equation}
which allows to retain definition \ref{eq-mh.mag.inst}
for the fluctuating magnetization.

When sites are updated successively on two checkerboard lattices 
defined by the parity of $i+j$, it is possible to show that the invariant 
measure of this system is the same as that of the Ising model
\cite{Sakaguchi92}:
\begin{equation}
\label{eq-inv.meas.saka}
P^{\rm eq}\left( \{ \sigma_{i,j} = m_{i,j} \} \right) \sim
\exp \; J \left\{ \sum_{i,j} m_{i,j} (m_{i-1,j} + m_{i+1,j} + 
m_{i,j-1} + m_{i,j+1} ) \right\}.
\end{equation}
Therefore, static exponents are known exactly (Onsager solution).
This was checked numerically in \cite{ising}, 
following the same procedure as that used for the Miller-Huse model.
Sakaguchi's model does indeed belong to the Ising universality class 
for static critical exponents.

However, nothing is known a priori on
dynamical properties, which we will investigate now 
along the same lines as before. One advantage of Sakaguchi's
model is that the critical point $J_c$ is known exactly. 
One source of error is thus eliminated.
One drawback is the following: the magnetization $M$ depends on 
site values at {\it two} consecutive time steps.
The method used previously in order to prepare initial conditions with
fixed small initial magnetization cannot work. 
Instead, we applied the following procedure: start from random initial 
conditions and evolve the system at high temperature 
(at low $J_{\rm init}$, the system is in the disordered phase with zero mean 
magnetization) until it reaches by itself the desired initial 
magnetization $m_0$. The system is then
quenched to the critical coupling.
Since the magnetization is calculated from discrete 
spins, there is no dispersion on the value of initial conditions
($m_0 = 2 K /L^2$, as before). 
Yet, this procedure is more costly numerically, which explains why,
in the following,
the obtained statistical accuracy is somewhat lower
than for the previous two models studied here. 
In practice, we use 
$J_{\rm init} = 0.1$, before quenching to $J_c = \ln(1+\sqrt{2})/2$.

\begin{figure}[thb]
\centerline{
\epsfxsize 10cm
\epsffile{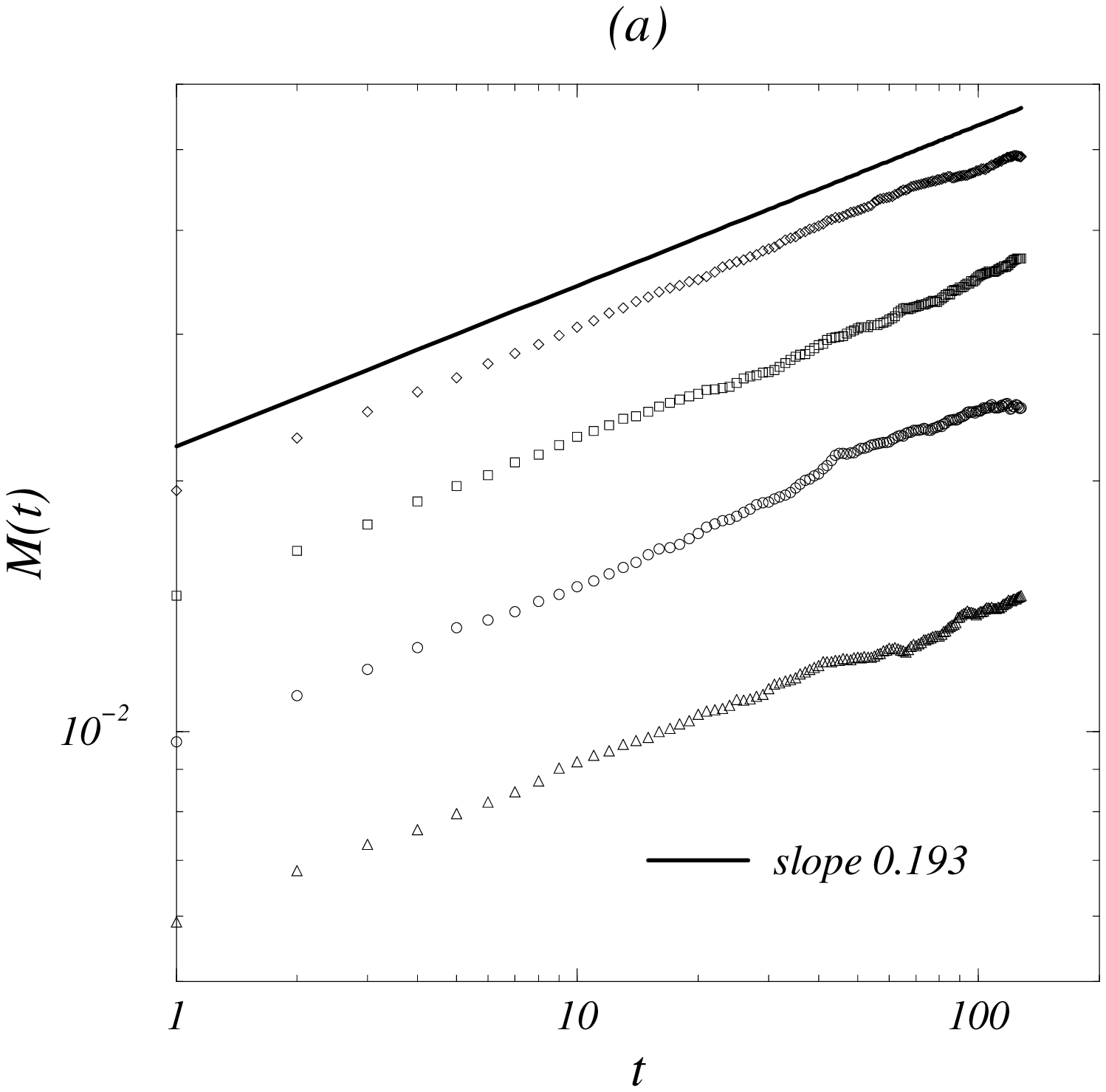}
\hspace{-1.5cm}
\epsfxsize 10cm
\epsffile{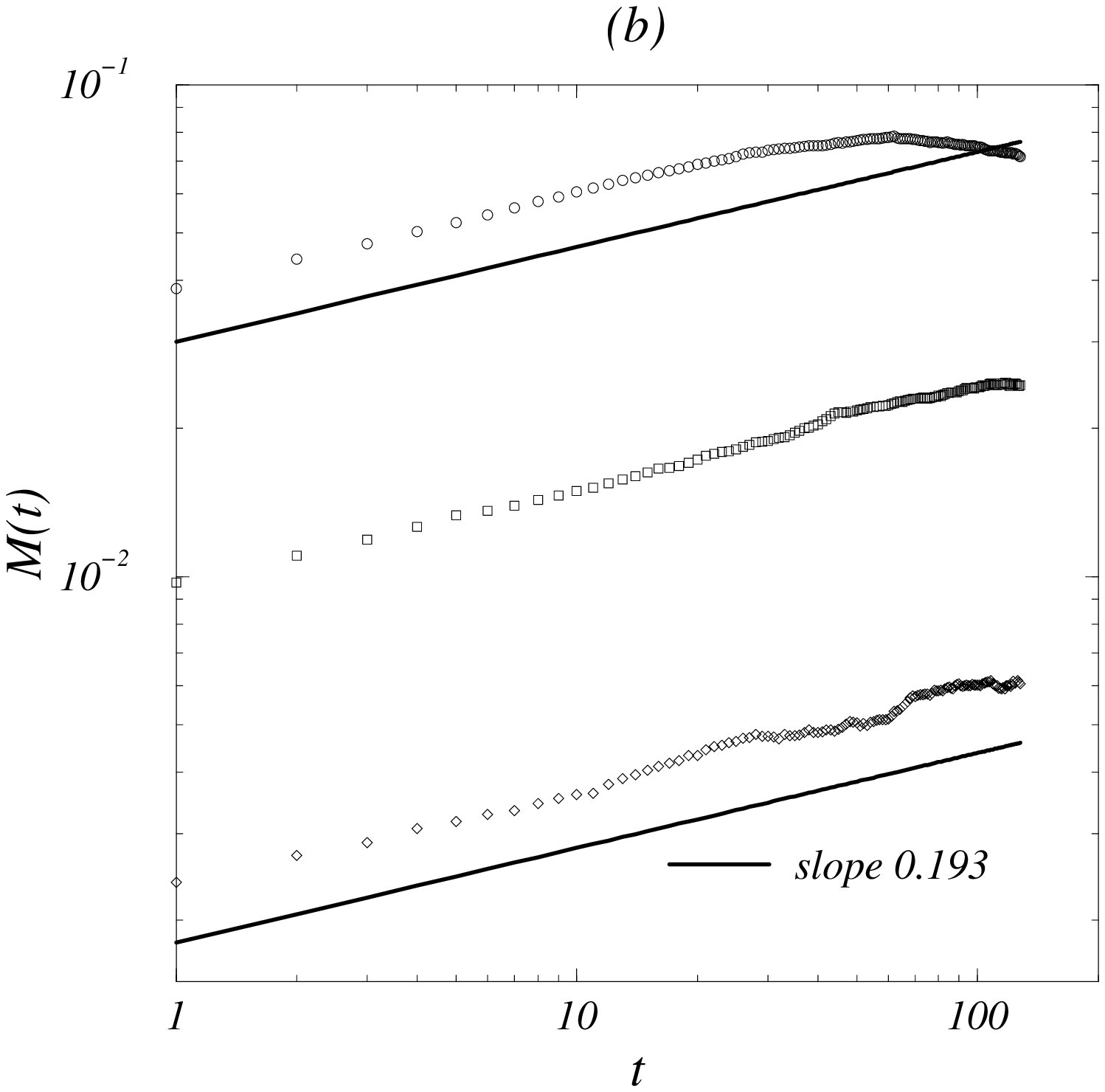}
}
\vspace{-0.5cm}
\caption{Measure of the dynamic critical exponent
$\theta'$ for Sakaguchi's model.
The magnetization $M(t)$, measured at the critical 
point $J = J_c$ (cf. text), is plotted vs. time $t$ 
in log-log scale. In both graphs,
the solid line corresponds to a 
slope equal to $\theta' = 0.193$ (Model A).
Graph $(a)$: the system size $L = 32$ is fixed, while
the initial magnetization $m_0 = 2 K /L^2$ 
varies between $m_0 = 7.8\;10^{-3}$ and $9.8\;10^{-3}$,
for values of $K = 4,6,8,10$ (from bottom to top).
The microscopic time $t_{\rm mic}$ is equal to zero. 
Scaling is thus observed over more than two decades.
Graph $(b)$: the initial condition $K = 4$ is fixed,
for system sizes $L = 16, 32, 64$ (from top to bottom).
The crossover time for the smallest size $L = 16$
is observable: $t_0 \sim 60$. Finite-size effects are 
negligible for $L = 32, 64$.
}
\label{fig-saka.theta}
\end{figure}

The protocol used is the same as before. As expected, this model also
exhibits a well-defined initial critical slip regime. 
In fact, the microscopic time $t_{\rm mic}$ turns out to be equal to $0$ 
(Fig.\ref{fig-saka.theta}). A possible explanation for this is that the
macroscopically-correlated domains present at $J_{\rm init}$
survive the quench to $j_c$. 
Impeccable scaling is observed for the initial
growth of the magnetization $M$. Reliable measurements are possible
only for $L = 32$ and $64$: cross-over to the nonlinear relaxation
regime occurs too soon ($t_0 \sim 60$) for $L = 16$,
and  $L = 128$ is too costly
to reach satisfactory statistical quality due to the preparation
of initial conditions. Ensemble average is done over $128000$ realizations
simulated during $T=128$ timesteps. 
From linear fits over the full interval 
$0 \le t \le 128$, we obtain $\theta' = 0.197(7)$ for $L = 32$ 
and $0.19(1)$ for $L = 64$ (Fig.\ref{fig-saka.theta}). 
Within error bars, no finite-size effects are present. 
Our estimate is:
\begin{equation}
\label{eq-saka-theta}
\theta'_{\rm Sakaguchi} = 0.194(14),
\end{equation}
in good agreement with results
obtained for Model A \cite{Grassberger,Okano}, and not consistent
with the value obtained for the Miller-Huse model.

\begin{figure}[thb]
\centerline{
\epsfxsize 10cm
\epsffile{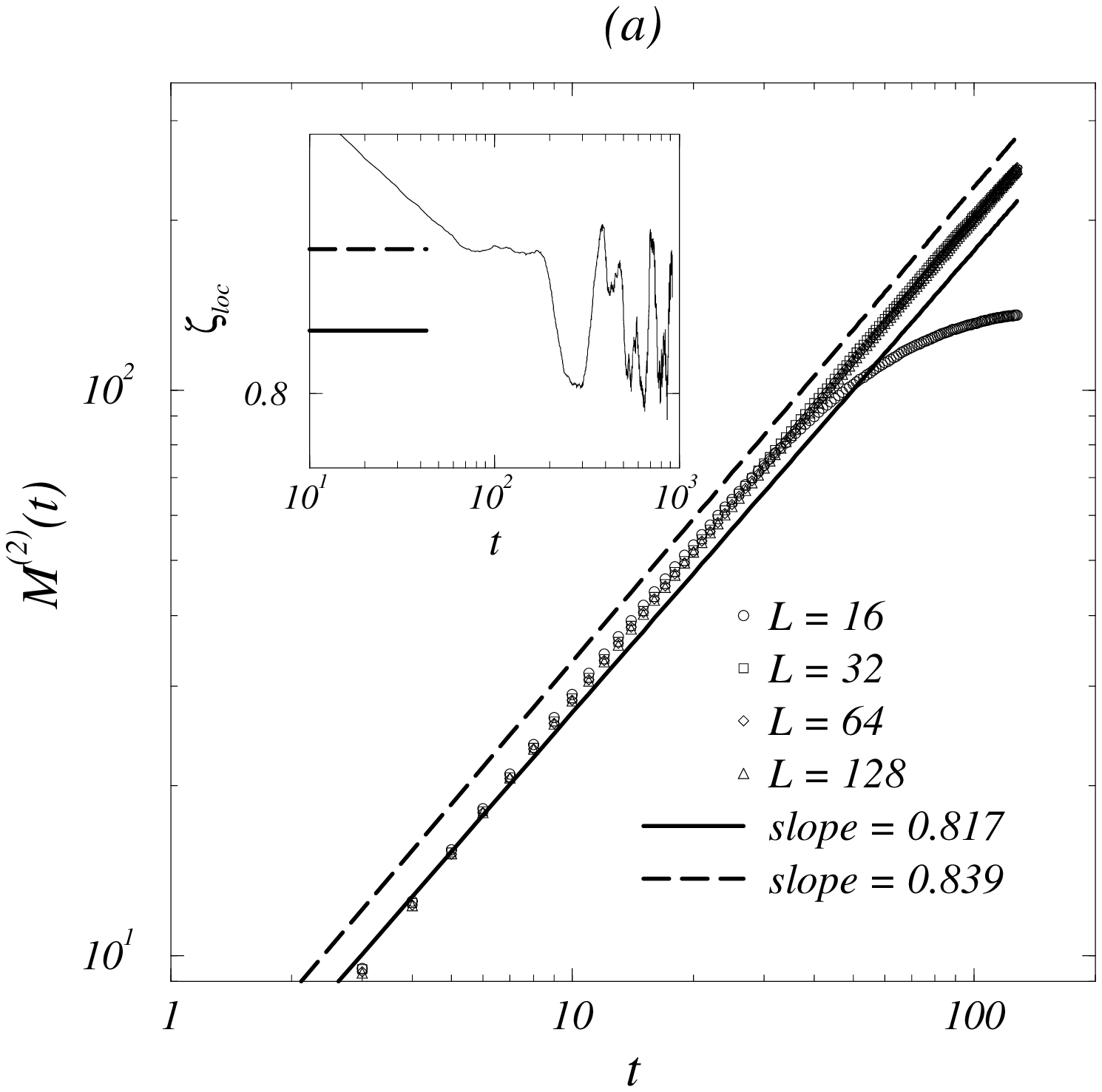}
\hspace{-1.5cm}
\epsfxsize 10cm
\epsffile{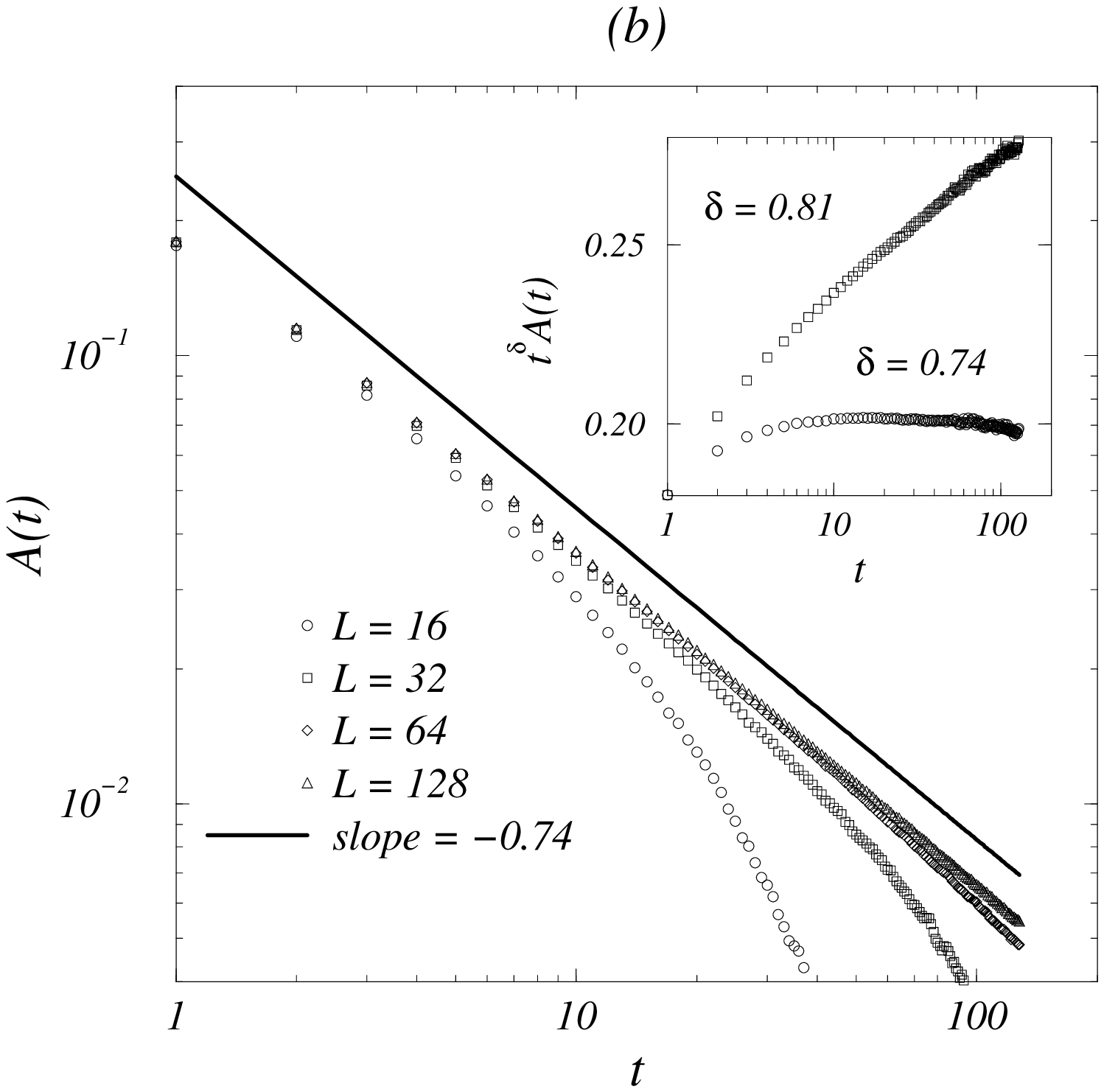}
}
\vspace{-0.5cm}
\caption{Measure of the dynamic critical exponent $z$
for Sakaguchi's model. The second moment
of the magnetization $M^{(2)}(t)$ (graph $(a)$) and the temporal
autocorrelation function $A(t)$  (graph $(b)$) are plotted
in log-log scale vs. time $t$, as measured at the 
(exact) critical point $J = J_c$. 
for zero initial magnetization $m_0 = 0$, up to
time $T = 128$. The system sizes 
considered are $L = 16, 32, 64, 128$, for $128000$ realizations.
Graph $(a)$: finite-size effects become negligible for $L \ge 32$.
The solid and dashed lines respectively correspond to slopes equal to 
$\zeta = 0.817$ (Model A) and $\zeta = 0.839$ (Miller-Huse model).
In the insert, we plot the local exponent $\zeta_{\rm loc}(t)$
computed over intervals $t_{\rm loc} = 100$, for $L=128$, $T = 1024$,
data averaged over $32000$ realizations.
Fluctuations of $\zeta_{\rm loc}(t)$ encompass both
values expected for Model A and the Miller-Huse model.
Graph $(b)$: the slope of the solid line is equal to 
$-\delta = -0.74$ (Model A).
The insert shows, for $L=128$, a log-lin plot of the ratio
$A(t)/t^{-\delta}$ vs. time $t$ for the
numerical values $\delta = 0.74$
(Model A, bottom curve) and $\delta = 0.81$
(Miller-Huse model, top curve). This suggests that the continuous
transition of Sakaguchi's model belongs to the
universality class of Model A for the exponent $\delta$.
}
\label{fig-saka.dyn}
\end{figure}

Again, measuring $z$ is more difficult. 
The scaling of $M^{(2)}$
starts late, with a fairly large microscopic time 
($t_{\rm mic} \sim 40$ for $L = 64$, $t_{\rm mic} \sim 70$ for $L = 128$), 
estimated as before from the evolution of local exponents.
As for non-zero initial magnetizations, ensemble average is performed
over $128000$ realizations, and the simulation time is $T=128$,
for system sizes up to $L = 128$. 
Exponents are thus estimated over less than one decade.
However, finite-size effects are negligible for
$L \ge 32$. We obtain:
\begin{equation}
\label{eq-saka-zeta}
\zeta_{\rm Sakaguchi} = 0.82(2)
\end{equation}
(note the large error bars). From the exact value 
$\beta/\nu = 1/8$, this leads to: $z = 2.13(5)$, 
roughly compatible with both
Model A and the Miller-Huse model. A longer simulation time
($T = 1024$), even for the largest system size ($L = 128$)
leads to qualitatively identical results, without any
quantitative improvement: the insert of Fig.~\ref{fig-saka.dyn}.a
shows that fluctuations of the local exponent $\zeta_{\rm loc}$,
computed over time intervals of duration $t_{\rm loc} = 100$,
include values of $\zeta$ expected for both Model A
and the Miller-Huse model.

The situation concerning the scaling behavior of the time 
autocorrelation function is somewhat more satisfactory,
despite (or thanks to) finite size effects:
large size behavior converges toward a value of $\delta$
which is necessarily smaller than the Miller-Huse 
exponent $\delta = 0.81(1)$.
In fact, data obtained for the largest size considered 
($L=128$, $T = 128$) is characterized by a scaling exponent
$\delta = d/z-\theta' \sim 0.75$, compatible with  
Model A value $\delta = 0.74(1)$ 
(cf. Fig.~\ref{fig-saka.dyn}.b). 
This finding is confirmed qualitatively by runs performed
for longer simulation times ($L = 128$, $T = 2048$):
the Miller-Huse value is not allowed
by the evolution of effective behavior (exponent) with system size.
Since we cannot cross-check with larger system size
due to numerical limitations, we do not provide error bars on 
our estimate:
\begin{equation}
\label{eq-saka-delta}
\delta_{\rm Sakaguchi} \sim 0.75.
\end{equation}

Our numerical results show clearly that the critical dynamics 
of Sakaguchi's model belongs neither to the universality
class of Miller-Huse model, nor to that of the sequentially
updated model studied in Sec.~\ref{sec-update-asyn}.
It is moreover likely that Sakaguchi's model belongs to the 
universality class of Model A, for both statics and dynamics: 
the checkerboard update of Sakaguchi's model may thus be analogous to
that of, say, Glauber dynamics of a spin system.

Our results may be summarized as follows
(see Table I for numerical values of all exponents
estimated so far): update is a relevant parameter for
dynamic critical exponents of Ising-like transitions of coupled chaotic
map systems. Not only models with static exponents outside the Ising
universality class (Miller-Huse) have also their dynamic exponents
different from those of Model A, but models within the Ising static class
(sequentially-updated Miller-Huse, Sakaguchi) may also have non-Ising
dynamic exponents (sequentially-updated Miller-Huse).
In some sense, this is not too surprising since update is already known
to be a relevant parameter for the dynamical exponents of spin systems in
the Ising static class. In the next section, we go a step further and 
investigate whether the universality found  at the static level
in \cite{ising} among synchronously-updated CMLs subsists at the dynamic
level.

\section{Non-universality within synchronously-updated models}
\label{sec-theta'}

In this section, we study two variants of the Miller-Huse model:
we consider first the case of a locally anisotropic coupling to three
neighbors, and then the case where the piecewise linear local map 
(\ref{eq-def.hm}) is replaced by a smooth map. The static exponents 
$\beta$, $\gamma$ and $\nu$ of
these variants were measured to be the same, within error bars, as those
of the Miller-Huse model.
Given the difficulties encountered above when trying to estimate the 
exponent $z$,
in this section we restrict ourselves to measurements of the 
exponent $\theta'$, via the early-time scaling of $M(t)$.

The three-neighbor variant of the Miller-Huse model was studied in 
\cite{ising} because of its particularly weak corrections to scaling.
Its anisotropic evolution rule reads:
\begin{equation}
\label{eq-sync.trois}
\left\{
\begin{array}{lcl}
x_{2i,j}^{t+1} &=& (1-3 g)\; f(x_{2i,j}^t) + g \; \left(
f(x_{2i-1,j}^{t}) +  f(x_{2i+1,j}^{t} +  f(x_{2i,j+1}^t) \right),\\
x_{2i+1,j}^{t+1} &=& (1-3 g)\; f(x_{2i+1,j}^t) + g \; \left(
f(x_{2i,j}^{t}) +  f(x_{2i+2,j}^{t} +  f(x_{2i+1,j-1}^t) \right),
\end{array}
\right.
\end{equation}
where $f$ is the original piecewise linear map (\ref{eq-def.hm}).
Each site is thus coupled to three of its nearest neighors:
sites belonging to even (resp. odd) columns of the lattice 
are coupled vertically to their northern (resp. southern) neighbor
only. Rule (\ref{eq-sync.trois}) is applied synchronously to all sites,
with periodic boundary conditions.
The local anisotropy introduced vanishes at large scales.
In this case, an Ising-like transition takes place for a critical
coupling $g_c=0.25118(4)$, estimated using Binder's method \cite{ising}.

The smooth map variant only differs from the Miller-Huse model by the choice
of the local function $f$, which remains a chaotic, odd map of the $[-1,1]$
interval for symmetry reasons, but now reads:
\begin{equation}
\label{eq-def.c4}
f(x) = 3 x - 4 x^3 \;.
\end{equation}
On general grounds, this smooth function, with its expanding and 
contracting parts, may be considered more ``generic'' than the
original piecewise linear map. An Ising-like transition also takes place
in this case, for a slightly smaller critical coupling than the Miller-Huse
model: $g_c = 0.17864(4)$ \cite{ising}.

\begin{figure}[thb]
\centerline{
\epsfxsize 10cm
\epsffile{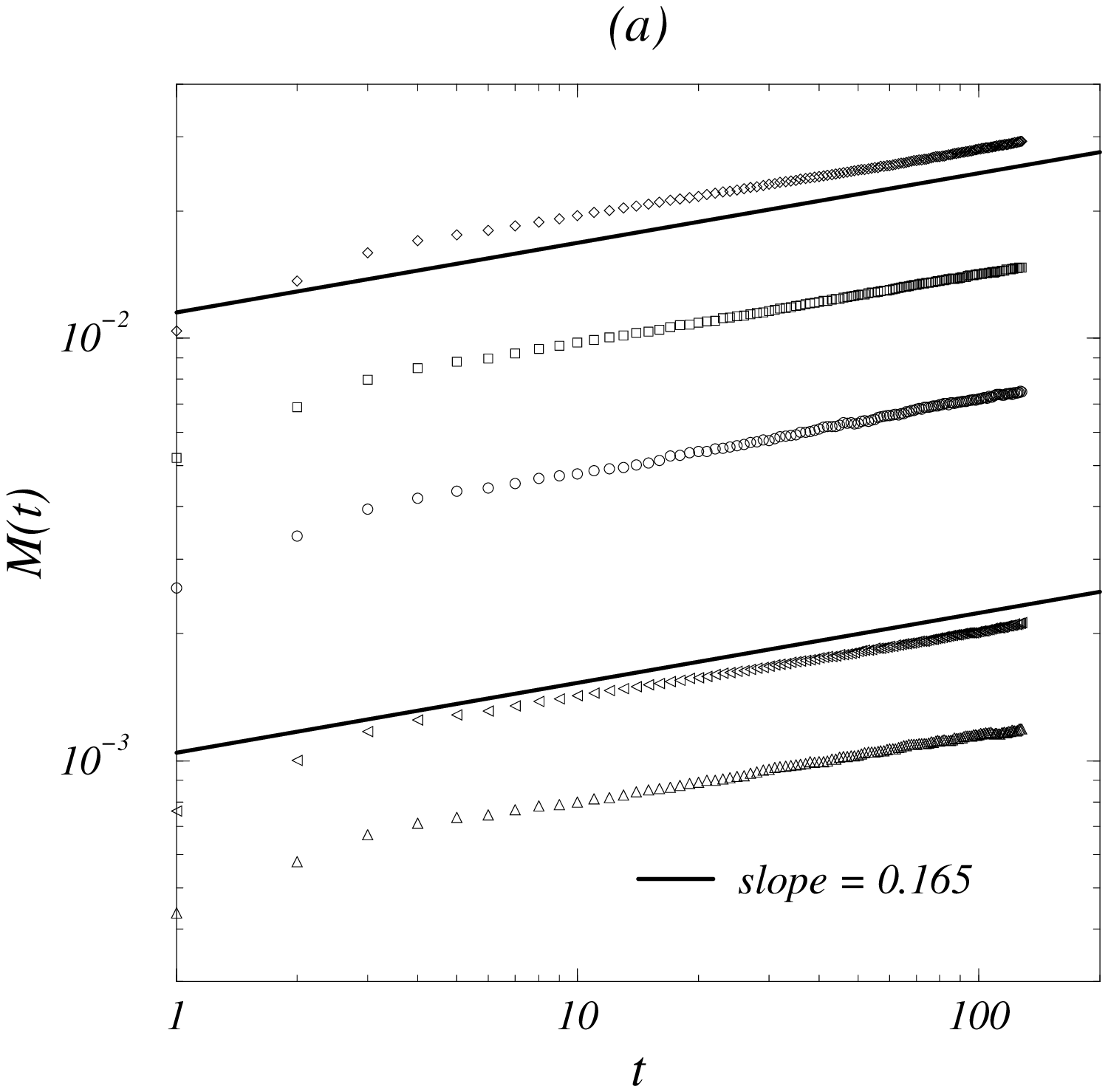}
\hspace{-1.5cm}
\epsfxsize 10cm
\epsffile{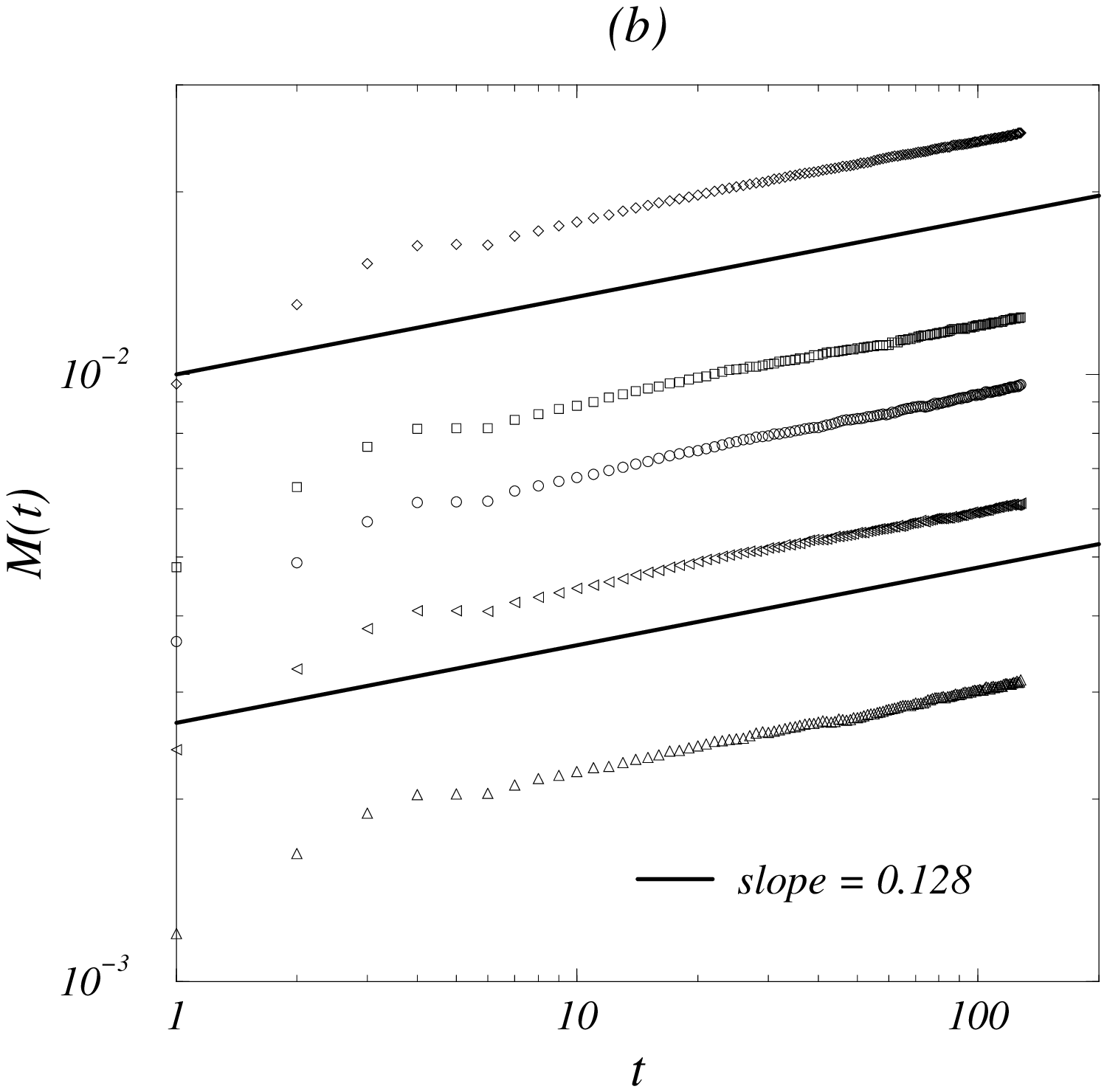}
}
\vspace{-0.5cm}
\caption{Measure of the dynamic critical exponent
$\theta'$ for two variants of the Miller-Huse model.
The magnetization $M(t)$, measured at the critical 
coupling $g_c$ (cf. text), is plotted vs. time $t$ 
in log-log scale. The system size is $L = 64$ 
(top three curves on each graph),
and $L=128$ (bottom two curves, shifted down for clarity). 
The initial magnetization is $m_0 = 2 K /L^2$. 
The microscopic time $t_{\rm mic}$ is equal to 5.
Graph $(a)$: anisotropic 
three-neighbor coupling  with original piecewise linear
local map (\protect\ref{eq-def.hm}) 
(model MH3). The solid lines correspond to 
slopes equal to $\theta' = 0.165$. From bottom to top: $L=128$, $K=8, 16$
and $L=64$, $K=2, 4, 8$.
Graph $(b)$: symmetric four-neighbor coupling with smooth cubic local map
(\protect\ref{eq-def.c4}) 
(model C4).
The solid lines correspond to slopes equal to $\theta' = 0.128$.
From bottom to top: $L=128$, $K=8, 16$ and $L=64$, $K=3, 4, 8$.
}
\label{fig-mh3c4.theta}
\end{figure}

For both variants, hereafter referred to as ``MH3'' (three-neighbor coupling)
and ``C4'' (smooth cubic map), the methodology used in Sec.~\ref{sec-mh} 
to determine exponent $\theta'$ for the original Miller-Huse model can
be applied. Initial conditions with fixed, small magnetization $m_0$ are 
prepared using the same procedure, and the phenomenology observed is the 
same. In both cases, the microscopic time is visually estimated to be
$t_{\rm mic} \sim 5$. After this time, a clear scaling behavior is observed
for $M(t)$ (Fig.~\ref{fig-mh3c4.theta}). 
Sizes $L=16$, 32, 64 and 128 were studied, with ensemble-averaging
over 512000 realizations for $L\le 64$ and 128000 for $L=128$. 
Finite-size effects are smaller than error bars for $L\ge 32$.
Statistical errors on $\theta'$ are estimated by comparing values obtained 
for different initial magnetization $m_0 = 2K/L^2$ with $K=2,4,8$ 
and coupling strengths $g$ within the uncertainty interval of $g_c$.
Our global (conservative) estimates are:
\begin{equation}
\theta'_{({\rm MH3})} = 0.165(8) \;\;\;\;{\rm and}\;\;\;\; 
\theta'_{({\rm C4})} = 0.128(6) \;.
\label{eq-mh3c4-cal}
\end{equation}
Not only these values are {\it not} compatible with 
each other, but both
are also at odds with the value found for the original Miller-Huse
model ($\theta' = 0.146(9)$). 

Numerical estimates of $\theta'$ 
obtained for all models we considered 
are gathered in Table II.
Except in the presence of finite-size
and finite-time effects not detectable in the experimental conditions
of our work, the above results lead to conclude that
CMLs of the Miller-Huse type exhibit different exponents for the dynamic
critical properties of their Ising-like transitions even though
they share the same static exponents and the same mode of update.

\section{Discussion}
\label{sec-disc}

The first conclusion about the series of numerical experiments
conducted in this work is that non-trivial early-time critical dynamics occurs
in phase transitions of far from equilibrium,
coupled map lattices. The microscopic time $t_{\rm mic}$ 
after which scaling sets in 
tends to be small for $M(t)$, generally of the same order of magnitude as for 
equilibrium systems. On the other hand, $t_{\rm mic}$ is typically
at least one order of magnitude larger for the evolution of second-order 
quantities such as $M^{(2)}(t)$ and $A(t)$. 
This is also in agreement with spin systems \cite{Okano}.
One would however like to understand why $t_{\rm mic}$ is always larger for
second-order quantities than for $M(t)$. Intuitively, it can be argued that
in order for the scaling regime of $M^{(2)}(t)$ and $A(t)$ to set in, 
the system first needs to generate 
macroscopically-correlated regions, i.e. $M(t)$
must already be in its scaling regime. Thus $t_{\rm mic}(M^{(2)}(t), A(t))
\ge t_{\rm mic}(M(t))$. Further, one may argue that 
macroscopically-correlated regions for second
order quantities can only establish themselves when the
magnetization $M(t)$ has changed ``significantly'',
say by a factor 2. Using $M(t) \sim t^{\theta'}$, one
obtains $t_{\rm mic}(M^{(2)}, A(t)) \sim 2^{1/{\theta'}}$.
This admittedly rough argument may explain 
semi-quantitatively why microscopic times
observed for the Miller-Huse models with sequential 
and synchronous update differ by two orders of 
magnitude: $t_{\rm mic}({\rm Seq.})/
t_{\rm mic}({\rm MH}) \sim 2^{(1/0.06)-(1/0.146)} \sim
2^{10}$, or $1024$.
In addition, finite-size corrections to scaling are also typically
much larger for second-order quantities than for $M(t)$.
A direct consequence of the observed values of $t_{\rm mic}$ and of the 
strength of finite-size corrections is that, in practice, 
the exponent $\theta'$ is much easier
to measure than the exponent $z$, our initial goal. Thus, although 
$z$ remains largely out of reach in some cases, 
the critical initial slip method
offers the advantage of accurate estimations of $\theta'$. 

The second conclusion of our work lies in the interpretation of the 
various dynamic exponent values we measured, which are summarized 
in Tables I and II.

First, and perhaps most importantly,
we confirm, at the level of dynamic critical properties, 
the results obtained in \cite{ising} at the static level: 
update is a relevant parameter for 
universality classes of Ising-like transitions. Synchronously-updated
coupled map models ---such as the Miller-Huse model and its variants MH3, C4---
have critical exponents different from those of Model A
both at the static and dynamic level. For the original Miller-Huse model,
our data is even conclusive for both exponents $\theta'$ and $z$.

Secondly, we extend to the transitions of coupled chaotic maps 
a result already known for equilibrium spin systems:
the static universality class of the Ising model breaks down at the
dynamic level. The sequentially-updated Miller-Huse model and the Sakaguchi
model, which are in the static Ising class, possess different dynamic 
exponents (in fact our data is unambiguous only for $\theta'$). 
The distinction between checkerboard and sequential update is
known to be irrelevant at equilibrium for both
static and dynamic critical exponents, including the exponent
$\theta'$ \cite{Grassberger,Okano}.
However, the same distinction becomes relevant
at the dynamic level for lattices of coupled chaotic maps.
An additional interesting point is that Sakaguchi's model seems to possess 
the dynamic exponents $\theta'$ and $z$ of Model A.

Thirdly, our data gathered for the sole exponent $\theta'$ indicates
the splitting of the ``non-equilibrium universality class'' for the static
critical properties of synchronously-updated coupled map systems
showing Ising-like transitions. 
Variants of the Miller-Huse model,
all shown in \cite{ising} to share the same static exponents,
exhibit different values of $\theta'$.  
Again, as for spin systems, universality classes are narrower
for dynamical critical exponents than for static ones.
Excluding Sakaguchi's model, whose control parameter 
is too markedly different to allow a meaningful 
comparison, one may notice that numerical estimates 
of $\theta'$ for the four other models depend monotonously
on the critical coupling $g_c$ (see Table II):
lower values of the coupling constant correspond to
a slower coarsening process. 
We believe this can be understood by considering the structure of the phase 
space of these systems which can be seen as a hierarchy of repellors.
During coarsening, this hierarchy is explored, and thus its scaling properties
--- which probably depend on $g$ --- must be related to the scaling behavior
which defines $\theta'$. Of course, the above picture will 
need to be substantiated in the future,
but it already suggests a link between the structure of the phase 
space (and hence the details of the dynamics) and exponent $\theta'$. 

In agreement with our findings, which confer a sort of
``maximal'' non-universality to $\theta'$, the tentative picture sketched
above,   invites comments on the relative status of $z$ and
$\theta'$. 
Unfortunately, our attempts at measuring $z$ 
(via $\zeta$ and $\delta$) have largely failed,
mainly because of large values of $t_{\rm mic}$.
However, data obtained for the variants C4 and MH3 of the Miller-Huse model
are compatible with the idea of a greater universality for $z$ than for
$\theta'$. In particular, they suggest that variant C4 might possess the same
$z$ value as the original Miller-Huse model.
Thus, we would like to suggest that $\theta'$ may be considered 
as a quantity highly dependent on the details of the dynamics in 
non-equilibrium situations, while $z$, in comparison, is a more global 
quantity. The independence of exponents $z$ and $\theta'$ is consistent
with this conjecture.

To summarize, our numerical experiments provide additional 
contradictory evidence 
to the conjecture of \cite{predictions} in the case of continuous
phase transitions of coupled chaotic map systems.
In other words, the qualitative 
features of scaling are correctly predicted by renormalization-group
methods (since, e.g., the scaling behavior predicted by \cite{Janssen89} 
is observed), but the same techniques fail to predict exponents 
quantitatively.
The above remarks, though, call for more detailed investigations of the 
respective role of the various dynamical exponents
involved in order to explain the origin of the non-universality 
reported here. In particular, the possibly different status of $z$
and $\theta'$ provides an interesting
starting point on which to base further research, at both the numerical and 
theoretical levels. Further, one would also like to know whether
the various persistence exponents defined and studied recently 
for spin systems \cite{persistence} ---and particularly the ``global
persistence'' exponent defined in \cite{global-per}--- 
also pertain to continuous
phase transitions of chaotic coupled map lattices, and if so,
whether their numerical value also depends on the fine details 
of the coarsening process following uncorrelated initial conditions,
similarly to the dynamic exponent $\theta'$. This is the subject of
ongoing research.

\acknowledgements
P.~M. acknowledges financial support from the European 
Union Science and Technology Fellowship 
Programme in Japan, and would like to thank 
the nonlinear physics group at Kyoto University
for its warm hospitality. Computer simulations were performed 
on the CRAY-T3D supercomputer at C.E.A.-Centre d'Etudes de Grenoble.

\newpage

\begin{table}
\label{tab-uni}
\begin{center}
\begin{tabular}{ccccc}
		& Model A	& Miller-Huse	& Sequential & Sakaguchi \\ \hline
Critical point	& $\ln(1+\sqrt{2})/2$ 	& $0.20534(2)$	&  $0.11255(5)$	& $\ln(1+\sqrt{2})/2$    \\ 
$\theta'$	& $0.193(5)$ 	& $0.146(9)$	& $0.06(1)$  	& $0.194(14)$ \\ \hline
$\zeta = (d-2 \beta/\nu)/z$	& $0.808(6)$ 	& $0.839(3)$	& $0.83(6)$  	& $0.82(2)$ \\
$z$		& $2.165(15)$ 	& $2.07(2)$	& $2.12(15)$   	& $2.13(5)$  \\ \hline
$\delta = d/z - \theta'$ & $0.74(1)$ & No estimate & $0.78(8)$ & $\sim 0.75$
\end{tabular}
\end{center}
\caption{Update is a relevant parameter for dynamic critical exponents:
a summary of numerical estimates of the dynamical exponents $\theta'$,
$\zeta$, $z$ (obtained from $\zeta$) and $\delta$, for three models
with synchronous (Miller-Huse model, Sec.~III),
sequential (Sec.~IV A), and checkerboard (Sakaguchi's model,
Sec.~IV B) update.
The exponent values given for Model A are discussed in Sec.~II.
Number(s) in brackets correspond to the uncertainty on the
last(s) digit(s), e.g. $0.193(5)$ means $0.193 \pm 0.005$,
}
\end{table}

\begin{table}
\label{tab-non-uni}
\begin{center}
\begin{tabular}{ccccccc}
		& 2D Ising	& Sakaguchi & Sequential & C4 	& Miller-Huse	& MH3	  \\ \hline
Static universality class	& Ising	&  Ising & Ising & Synchronous	&  Synchronous	& Synchronous   \\ 
Critical point	& $\ln(1+\sqrt{2})/2$ & $\ln(1+\sqrt{2})/2$ & $0.11255(5)$ &
$0.17864(4)$	& $0.20534(2)$	& $0.25118(4)$ \\
$\theta'$	& $0.193(5)$ 	& $0.194(14)$ & $0.06(1)$ &  $0.128(6)$  & $0.146(9)$	& $0.165(8)$  
\end{tabular}
\end{center}
\caption{The universality observed for
the static properties of Ising-like transitions of 
lattices of coupled chaotic maps
does not hold for their dynamic critical properties: 
a summary of numerical estimates 
of the dynamical critical exponents $\theta'$.
Number(s) in brackets correspond to the uncertainty on the
last(s) digit(s), e.g. $0.193(5)$ means $0.193 \pm 0.005$,
}
\end{table}

\end{document}